\documentclass[aps,prb]{revtex4}
\usepackage[dvips]{graphicx}

\begin{document}
\title{
Simulation Study of the One-Dimensional Burridge-Knopoff Model of Earthquakes}

\author{Takahiro Mori and Hikaru Kawamura}
\affiliation{Department of Earth and Space Science, Faculty of Science,
Osaka University, Toyonaka 560-0043,
Japan}
\date{\today}
\begin{abstract}
Spatio-temporal correlations of the one-dimensional spring-block
 (Burridge-Knopoff) model of earthquakes are  extensively studied by
 means of numerical computer simulations. Particular attention is paid
 to clarifying how the statistical properties of earthquakes depend on
 the frictional and elastic properties of earthquake faults. It is found
 that, as the velocity-weakening tendency of the friction force gets
 weaker, the system tends to be more critical, while as the
 velocity-weakening tendency gets stronger the system tends to be more
 off-critical with enhanced features of a characteristic earthquake. The
 model exhibits several eminent precursory phenomena prior to the large
 event in its spatio-temporal correlations. Preceding the mainshock, the
 frequency of smaller events is gradually
enhanced, whereas, just before the mainshock, it is 
suppressed in a close vicinity of the epicenter of the upcoming event (the Mogi doughnut). 
The time scale of the onset of the doughnut-like quiescence depends on the
extent of the frictional instability.
Under certain conditions, preceding the mainshock, 
the apparent $B$-value of the magnitude distribution increases significantly.
The existence of such distinct precursory phenomena 
may open a way to the prediction of the time and the position of 
the upcoming large event. 
\end{abstract}
\maketitle

\section{Introduction}

Earthquake is a stick-slip frictional instability of a fault
driven by steady motions of tectonic plates [\textit{Scholz},
1990;\textit{Scholz}, 1998].
While earthquakes are obviously  complex phenomena, 
certain empirical laws have been known concerning their statistical properties,
{\it e.g.\/}, the Gutenberg-Richter (GR) law for the magnitude distribution of
earthquakes, or
the Omori law for the time evolution of the frequency of 
aftershocks [\textit{Scholz}, 1990]. 
These laws are basically
of statistical nature, becoming eminent only after analyzing large number of
events. 

Since earthquakes could be regarded as a stick-slip frictional 
instability of
a pre-existing fault, the statistical properties of earthquakes 
should be governed  by the physical law of 
rock friction [\textit{Scholz},
1990;\textit{Scholz}, 1998]. 
One might naturally ask: How the statistical properties of earthquakes 
depend on the material properties characterizing earthquake faults, 
{\it e.g.\/},
the elastic properties of the crust or the frictional properties of 
the fault, {\it etc\/}. 
Answering such questions 
undoubtedly would give us
valuable information in understanding the true nature of  earthquakes.
Furthermore, the full knowledge of spatio-temporal correlations of earthquakes, particularly 
those associated with the precursory phenomena of large events, might
ultimately open up a door to the statistical prediction of large events.

Meanwhile,  a systematic field study
of the material-parameter
dependence of the statistical properties of
earthquakes meets serious difficulties, partly because one has
to average over large number of events in obtaining the  
statistically reliable properties of earthquakes, but also because 
it is difficult to get precise knowledge of, or even to control, various 
material
parameters characterizing real earthquake faults.

In numerical simulation study of earthquakes, on the other hand,
these difficulties often become minor ones. 
In the past,  earthquake models of various
levels of simplifications  have been proposed in geophysics
and statistical physics, and their statical properties have been extensively
studied mainly by means of numerical computer simulations.
One of the most standard one is the so-called 
spring-block model originally proposed by Burridge and Knopoff 
(Burridge-Knopoff model) [\textit{Burridge and Knopoff}, 1967]. 
In this model, an earthquake fault is
simulated by an assembly of blocks, each of which is connected via 
the elastic springs to the neighboring blocks and to 
the moving plate. 
All blocks are subject to the 
friction force, the source of the nonlinearity in the
model, which eventually realizes an earthquake-like frictional instability. 
The model contains several parameters representing, {\it e.g.\/},  
the elastic properties of the crust and the
frictional properties of faults. 

Of course, the space discretization in the
form of blocks is an approximation to the continuum crust.
Although the true meaning and the possible
effect of the 
block discretization needs fuller examination, its effect is expected
to be less for a long-scale behavior associated with large events. 
In spite of this limitation, the spring-block model has served 
as a very useful reference model of earthquakes for years.

Carlson, Langer and collaborators
performed a pioneering study of 
the statistical properties of 
earthquakes [\textit{Carlson and Langer}, 1989a;\textit{Carlson and
Langer}, 1989b;\textit{Carlson et al.}, 1991;\textit{Carlson},
1991a;\textit{Carlson}, 1991b;\textit{Carlson et al.}, 1994], 
based on the spring-block model.
These authors paid particular attention to
the magnitude distribution of earthquake events, and 
examined its dependence on the friction parameter characterizing the 
nonlinear stick-slip dynamics of the model. It was observed that,
while smaller events persistently obeyed the GR law, {\it i.e.\/},
staying critical or near-critical, 
larger events exhibited a significant 
deviation from the GR law, being off-critical 
or ``characteristic''[\textit{Carlson and Langer},
1989a;\textit{Carlson and Langer}, 1989b;\textit{Carlson
et al.}, 1991;\textit{Carlson}, 1991a;\textit{Carlson}, 1991b;\textit{Schmittbuhl et al.}, 1996].  
Shaw, Carlson and Langer studied the same model by
examining the spatio-temporal patterns
of seismic events preceding large events, observing that
the seismic activity accelerates as the large event approaches
[\textit{Shaw et al.}, 1992].
Since then, the spring-block model has been extended in several ways, 
{\it e.g.\/}, taking account of the effect of the viscosity 
[\textit{Myers and Langer}, 1993;\textit{Shaw}, 1994;\textit{De and Ananthakrisna}, 2004], or modifying the form of the friction force
[\textit{Myers and Langer}, 1993;\textit{De and Ananthakrisna}, 2004]. 
The study of statistical properties of earthquakes was promoted in  early 
nineties,
inspired by the  work by P. Bak and collaborators
who proposed the concept of 
``self-organized criticality (SOC)'' [\textit{Bak et al.}, 1987;\textit{Bak and
Tang}, 1989].
According to this view, the Earth's crust is always in the critical state
which is self-generated dynamically. 
The SOC idea gives a natural
explanation of the scale-invariant power-law behaviors frequently observed in
earthquakes, including the GR law and the Omori law. 

The SOC idea was developed mainly on the basis of  
the cellular-automaton versions of the earthquake 
model [\textit{Bak et al.}, 1987;\textit{Bak and Tang},
1989;\textit{Nakanishi}, 1990;\textit{Ito and Matsuzaki},
1990;\textit{Brown et al.}, 1991;\textit{Olami et al.},
1992;\textit{Hergarten et al.}, 2000;\textit{Hainzl et al.},
1999;\textit{Hainzl et al.}, 2000;\textit{Helmstetter et al.},
2002;\textit{Helmstetter et al.}, 2004].
The statistical properties of these cellular-automaton models
were also studied quite extensively, often
interpreted within the SOC framework. These models apparently reproduce
several fundamental features of earthquakes such as the GR law, the Omori law of aftershocks, the existence
of foreshocks, {\it etc\/}. 
Although many of these cellular-automaton models were  originally
introduced to mimic the
spring-block 
model, 
their statistical properties
are not always identical with the original spring-block model.
Furthermore, as compared with the spring-block model,
the cellular-automaton models are much more simplified so that
the model does not have  enough room to represent various material 
properties of the earthquake fault in a physically appealing way.
Thus,  in the cellular-automaton models, the connection 
to the material parameters of real earthquake faults 
is rather obscure.

In the present paper, we  perform further extensive investigation of
spatio-temporal correlations of earthquakes based 
on the Burridge-Knopoff (BK) model.
Our main  goal is to clarify how the statistical properties
of earthquakes depend on the frictional and elastic properties of earthquake
faults. 
As compared with the cellular-automaton
models, the spring-block model has an advantage that the dependence on the
material parameters, including  the frictional and elastic properties, 
are more explicit.
We simulate  the one-dimensional (1D) version of the 
BK model. The statistical properties of the model, particularly 
its spatio-temporal correlations, are studied here by 
extensive numerical simulations. A preliminary report of the simulation was
given in [\textit{Mori and Kawamura}, 2005].

As already mentioned, numerical model simulation is a quite useful tool
for our purposes:
First, generating huge number of events required to attain the 
high precision for discussing the statistical properties of
earthquakes is
easy to achieve in numerical simulations whereas it
is often difficult to achieve in real earthquakes, especially
for larger ones.
Second,  various material parameters characterizing earthquake faults
is extremely 
difficult to control in real faults,  whereas it is easy to control
in model simulations.


Real faults are of course not 1D. Although it is 
desirable to study the 2D and even the 3D models,
we concentrate ourselves on the 1D model here. 
This is because the 1D model is much easier to
simulate than the higher-dimensional counterparts, which enables
us to obtain better statistics  necessary to get reliable information
about spatiotemporal correlations of the model. Most importantly,
it enables us to study large enough systems and systematically examine 
the finite-size effect. The extension to 2D might be done in two ways: In one,
the 2D model is regarded to represent the fault plane itself [\textit{Carlson}, 1991b], while in 
the other the second direction of the model is taken to be orthogonal to the
fault plane [\textit{Myers et al.}, 1996]. 
These numerical studies so far made on such 2D BK models,
mainly concerning their magnitude distribution, suggested that the
basic features of the 2D BK model
might not drastically be different from those of the 1D model.
Indeed, our preliminary study on the 2D model, of the first type mentioned
above, also supports this observation.


Although the model
studied here, {\it i.e.\/}, the 1D BK model
is not new and has been studied by previous authors, we simulate the model by
more extensive simulations than the previous ones, 
and investigate the statistical properties of the model
through various spatio-temporal correlation functions. 
As a result,
our present simulations have revealed several new interesting features of 
the model which have hitherto uncovered: These include, 
i) the doughnut-like quiescence occurring just before the mainshock, 
ii) a significant change of the $B$-value observed prior to the
mainshock, and iii) complex behaviors of the recurrence time of large
events including the possible twin and unilateral character of 
large events, {\it etc\/}. 

Generally, we have found that as the velocity-weakening tendency of the 
friction force gets
weaker, the system tends to be more critical, while as the
velocity-weakening tendency gets stronger, the system tends to be more 
off-critical with enhanced features of a characteristic earthquake.
Periodic feature of large events is eminent when
the friction force exhibits a strong frictional instability, whereas,
when the friction force exhibits a weak frictional instability, 
large events often occur as twin and/or unilateral events.
The model turns out to exhibit several eminent precursory
phenomena prior to the large event in its spatio-temporal correlations.
Preceding the mainshock, the frequency of smaller 
events is gradually
enhanced, whereas, just before the mainshock, it is 
suppressed in a close vicinity of the epicenter of the upcoming mainshock
(the Mogi doughnut) [\textit{Mogi}, 1969;\textit{Mogi}, 1979]. 
The time scale of the onset of the doughnut-like quiescence depends on the
extent of the frictional instability.
Under certain conditions, preceding the mainshock, 
the apparent $B$-value of the magnitude distribution increases significantly.
This increase of the $B$-value is more eminent 
when the friction force exhibits a strong frictional instability.
The existence of such distinct precursory phenomena 
may open a way to the prediction of the time and the position of 
the upcoming large event. 

The present paper is organized as follows. In \S 2, we introduce the model
and explain some of the details of our numerical simulation. 
The results of our simulations are presented in \S 3. Magnitude distribution
function, earthquake recurrence-time distribution and various types of 
spatio-temporal correlation functions of earthquake events are calculated
and analyzed. In search of the possible precursory phenomena,  
particular attention is paid to the time development of 
the spatial correlation function of earthquake events and 
of the magnitude distribution function
prior to large events.
Finally, section 4 is devoted to summary and discussion of our results.

\section{The model and the method}

The model we deal with  is the one-dimensional Burridge-Knopoff (BK) model.
It consists of a 1D array of $N$ identical blocks, 
which are mutually connected with the two neighboring 
blocks via the 
elastic springs of the 
elastic constant $k_c$, and are also connected to 
the moving plate
via the springs of the elastic constant $k_p$.
All blocks are subject to the 
friction force $\Phi$, which is the only source of nonlinearity in the
model. The equation of motion for the $i$-th block can be
written as
\begin{equation}
m \ddot U_i=k_p (\nu ' t'-U_i) + k_c (U_{i+1}-2U_i+U_{i-1})-\Phi (\dot U_i),
\end{equation}
where $t'$ is the time, $U_i$ is the displacement of the 
$i$-th block and
$\nu '$ is the loading rate 
representing the speed of the plate. 

In order to make the equation
dimensionless, we measure the time $t'$ in units of the characteristic 
frequency $\omega =\sqrt{k_p/m}$ and the displacement $U_i$ in units of
the length $L=\Phi(0)/k_p$, $\Phi(0)$ being the static friction. Then,
the equation of motion can be written  in the dimensionless form as
\begin{equation}
\ddot u_i=\nu t-u_i+l^2(u_{i+1}-2u_i+u_{i-1})-\phi (\dot u_i),
\end{equation}
where $t=t'\omega $ is the dimensionless time, 
$u_i\equiv U_i/L$ is the dimensionless displacement of the 
$i$-th block, 
$l \equiv \sqrt{k_c/k_p}$ is the dimensionless stiffness parameter, 
$\nu =\nu '/(L\omega)$ is the dimensionless loading rate, and  
$\phi(\dot u_i) \equiv \Phi(\dot U_i)/\Phi(0)$ is the dimensionless friction 
force.

In order for the model to exhibit a 
dynamical instability corresponding to an earthquake, it is essential 
that the friction force $\phi$ possesses a frictional {\it weakening\/}
property, {\it i.e.\/},
the friction should become weaker as the block slides.
Here, as the form of the friction force, 
we assume the form used by \textit{Carlson et al.} [1991], 
which represents the
velocity-weakening friction force;
\begin{equation}
\phi(\dot u) = \left\{ 
             \begin{array}{ll} 
             (-\infty, 1],  & \ \ \ \ {\rm for}\ \  \dot u_i\leq 0, \\ 
              \frac{1-\sigma}{1+2\alpha \dot u_i/(1-\sigma )}, &
             \ \ \ \ {\rm for}\ \  \dot u_i>0, 
             \end{array}
\right.
\end{equation}
where its maximum value corresponding to the static friction
has been normalized to unity. As noted above, this normalization 
condition $\phi(\dot u=0)=1$ has been utilized to set the length unit $L$.
The back-slip is inhibited by imposing an
infinitely large friction for $\dot u_i<0$, {\it i.e.\/}, 
$\phi(\dot u<0)=-\infty $. 

The friction force is characterized by the two parameters, $\sigma$ and 
$\alpha$. The former, $\sigma$, 
introduced by \textit{Carlson et al.} [1991] as a technical device,  
represents an instantaneous drop of the friction force
at the onset of the slip, while the latter, $\alpha$, 
represents the rate of the friction force getting weaker
on increasing the sliding velocity. In the present simulation, 
we regard $\sigma$ to be small, and fix
$\sigma =0.01$. 

We note that there have been several other proposals for the law of rock
friction, {\it e.g.\/}, the slip-weakening friction 
force [\textit{Scholz}, 1990;\textit{Shaw},
1995;\textit{Myers et al.}, 1996;\textit{Scholz}, 1998] 
or the rate- and state-dependent 
friction force [\textit{Dietrich}, 1979;\textit{Ruina}, 1983;\textit{Scholz}, 1990;\textit{Scholz}, 1998]. 
We assume here the simplest version of the velocity-weakening 
friction force given above. 
 
We also assume the loading rate $\nu$ to be infinitesimally small, and put 
$\nu=0$ during an earthquake event, a very good approximation 
for real faults[\textit{Carlson et al.}, 1991]. Taking this limit ensures that the interval
time during successive
earthquake events can be measured in units of $\nu^{-1}$ 
irrespective of 
particular values of $\nu$. Taking the $\nu \rightarrow 0$ limit
also ensures that, during an ongoing event,
no other event takes place at a distant 
place, independently of this ongoing event. 

Then, the model possesses one more
parameter, a dimensionless stiffness parameter  $l$, which describes
the elastic property of the crust.

We solve the equation of motion (2) 
by using the Runge-Kutta method of the fourth 
order. The width of the time discretization $\Delta t$ is taken to be
$\Delta t\nu =10^{-6}$. We have checked that the statistical properties 
given below are unchanged
even if we take the smaller $\Delta t$. Total number of $10^7$  events
are generated in each run, 
which are used to perform various averagings. In calculating the observables,
initial $10^4$ events are 
discarded as transients.


In order to eliminate the possible finite-size effects, the total
number of blocks are taken to be large, $N=800\sim 6,400$, with
periodic boundary conditions. In the case of smaller $l$, say, 
$l=3$,  even the largest 
event involves the number of blocks less than $N=800$.
By contrast, in the case of larger $l$, the finite-size effect
becomes more significant. In such cases, we simulate the system as
large as $N=6,400$ so that the results are free from finite-size effects.


We study the properties of the model with varying the frictional parameter
$\alpha$ and the elastic parameter $l$. Particular attention is paid to
the dependence on the parameter 
$\alpha$, since the parameter $\alpha$, which represents the
extent of the frictional weakening,  
turns out to affect the result most significantly.

\section{The results}

In this section, we show the results of our numerical simulations
for several observables, {\it i.e.\/}, magnitude distribution
of events, mean displacement of blocks at each large event, local 
recurrence-time distribution of large events, 
global recurrence-time distribution of large events, magnitude
correlations between successive large events, 
time correlations of events 
associated with the mainshock, 
spatial correlations of events before the mainshock, spatial
correlations of events after the mainshock, 
time-resolved magnitude distribution of
events before and after the mainshock, {\it etc\/}. 
Below, we show these results consecutively.

\subsection{Magnitude distribution}

\begin{figure}[ht]
\begin{center}
\includegraphics[scale=0.65]{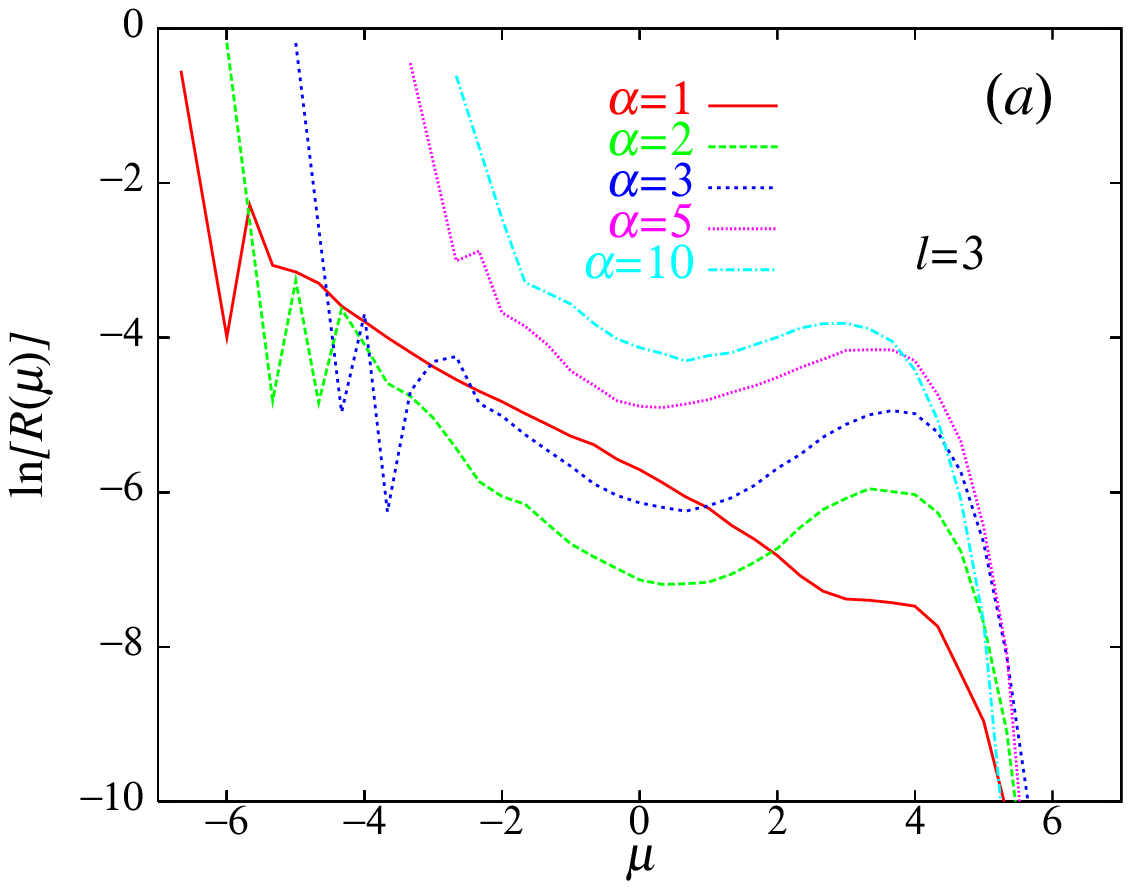}
\includegraphics[scale=0.65]{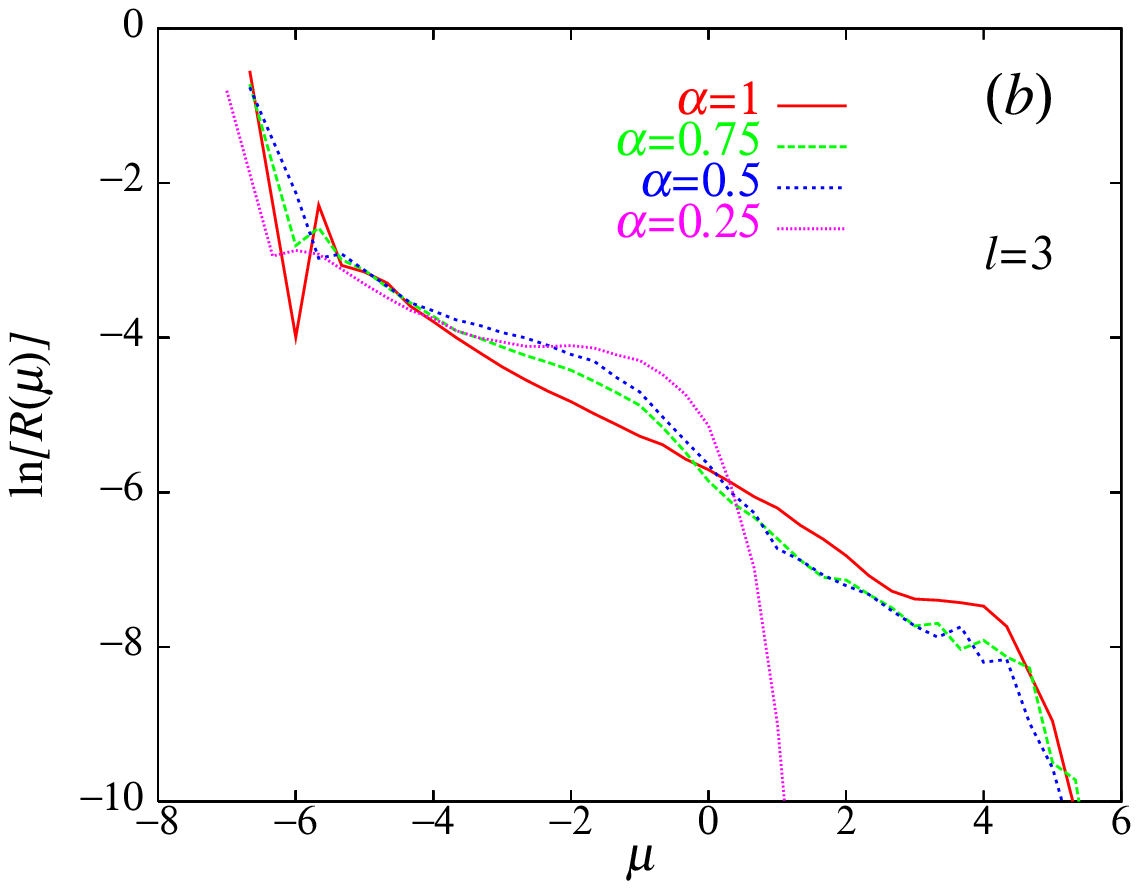}
\end{center}
\caption{
Magnitude distribution of earthquake events for various values of $\alpha$; (a)
for larger $\alpha =1,2,3,5$ and 10, and (b) for smaller 
$\alpha =0.25, 0.5, 0.75$ 
and 1. The parameter $l$ is fixed to be $l=3$. The system size is $N=800$.
}
\end{figure}

In Fig.1 we show the magnitude
distribution $R(\mu)$ of earthquake events, 
where $R(\mu){\rm d}\mu$ represents the rate of events with their
magnitudes in the range [$\mu, \mu +{\rm d}\mu$]. The parameter $\alpha$ 
is varied in the range
$0.25 \leq \alpha\leq 10$, while the parameters $l$ 
 is fixed to $l=3$.

The magnitude of an event, $\mu$, 
is defined as the logarithm of the moment $M_0$, {\it i.e.\/}, 
\begin{equation}
\mu=\ln M_0, \ \ \ M_0=\sum_i \Delta u_i,
\end{equation}
where $\Delta u_i$ is the 
total displacement of the $i$-th block during a given 
event and the sum is taken over all blocks involved in the event
[\textit{Carlson et al.}, 1991].

As can be seen from Fig.1, the form of the calculated $R(\mu)$ depends on the
$\alpha$-value considerably.
The data  for  $\alpha =1$ lie on a straight line fairly well, apparently
satisfying the GR law. The values of the exponent $B$ describing 
the power-law behavior, 
$\propto 10^{-B}$, is estimated to be $B\simeq 0.50$.

As can be seen from Fig.1(a), 
the data for larger $\alpha$, {\it i.e.\/}, the ones for $\alpha \geq 2$
deviate
from the GR law at larger magnitudes, exhibiting a pronounced peak structure, 
while the power-law feature still remains for smaller magnitudes.
These features of the magnitude distribution are consistent with the
earlier observation of Carlson and Langer [\textit{Carlson and
Langer},1989a;\textit{Carlson and Langer} 1989b;\textit{Carlson et al.}, 1991]. It means that,
while smaller events 
exhibit self-similar critical properties,  larger events tend to exhibit
off-critical or characteristic properties, much more so as
the velocity-weakening tendency of the friction  is increased.

The observed peak structure 
gives us a criterion to distinguish large and small events. Below,
we regard events with their magnitudes $\mu$ greater than $\mu_c=3$
as large events, $\mu_c=3$ being 
close to the peak position of the magnitude distribution of Fig.1. 
In an earthquake with  $\mu=3$,
the mean number of moving blocks are about 76 ($\alpha =1$) and 60 
($\alpha =2,3$).

By contrast, as can be seen from Fig.1(b), the data for smaller $\alpha <1$
exhibit considerably different behaviors from those for $\alpha >1$.
Large events are suppressed here. For $\alpha =0.25$, in particular, 
all events consist almost exclusively of small events only. 
This result might be consistent with the earlier observation by 
\textit{Carlson and Langer}, [1989a],
which suggested that 
the smaller value of $\alpha <1$ tended to cause 
a creeping-like behavior without a large event. 
In particular, Vasconcelos showed that a single block system 
exhibited a ``first-order
transition'' at $\alpha =0.5$ from a stick-slip to 
a creep [\textit{Vasconcelos}, 
1996], whereas this
discontinuous transition becomes apparently continuous in many-block system
[\textit{Vieira et al.}, 
1993; \textit{Clancy and Corcoran}, 2005].
Since  we are mostly
interested in large seismic events  in the present paper, 
we concentrate  in the following on the parameter range
$\alpha \geq 1$. 
We mainly study the cases $\alpha =1,2$ and 3 since
further increase of $\alpha >3$ does not 
change the results qualitatively, as can be seen from Fig.1(a). 

It should be noticed that the magnitude 
distribution exhibits somewhat irregular
behaviors at small magnitudes. Such an irregular behavior of $R(\mu)$ 
observed at small
magnitudes arises not due to the insufficient statistics of our simulation, 
but rather reflects the
intrinsic discreteness of the model, {\it i.e.\/}, 
the fact that the magnitudes of events involving only a few number of blocks
tend to be determined to some specific values.
This is an artifact originating 
from the discreteness of the BK model. Of course, at larger magnitudes where 
more blocks are involved, 
the discreteness of the model becomes increasingly irrelevant, and the smooth
distribution function arises.

\begin{figure}[ht]
\begin{center}
\includegraphics[scale=0.65]{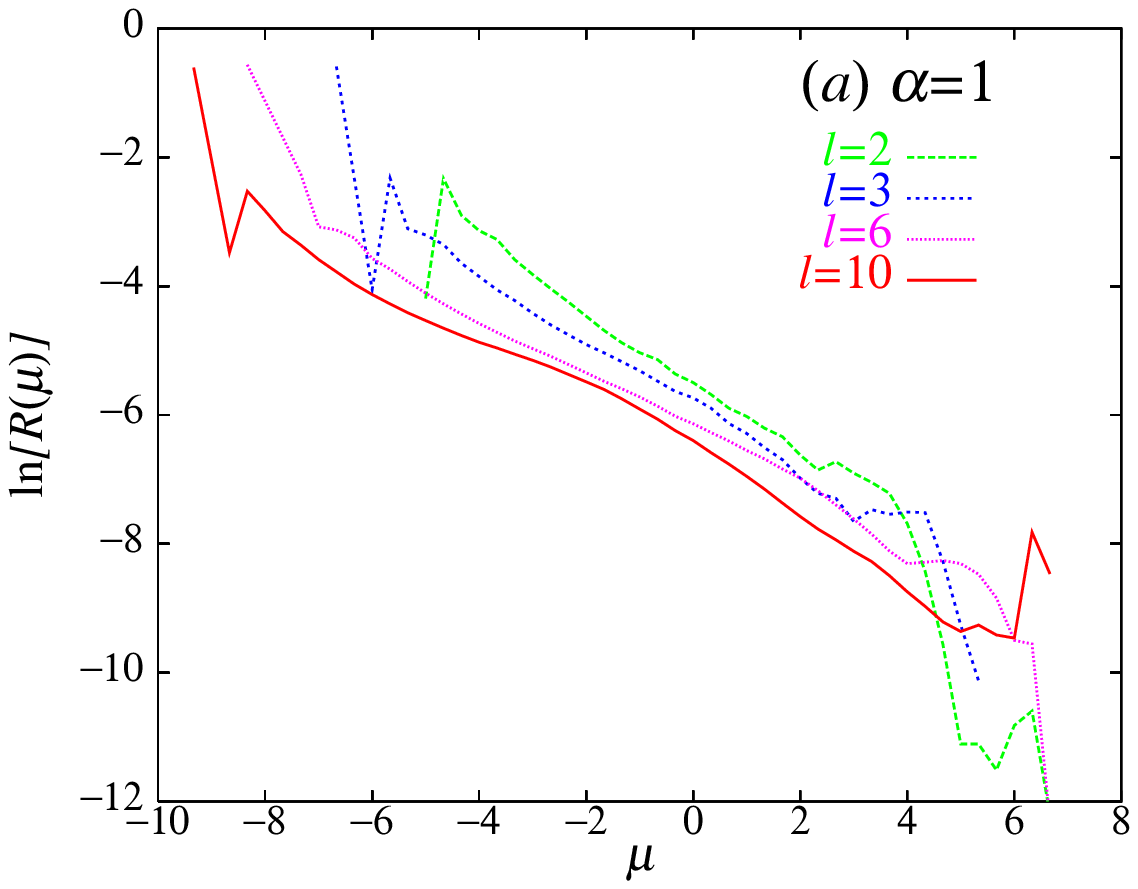}
\includegraphics[scale=0.65]{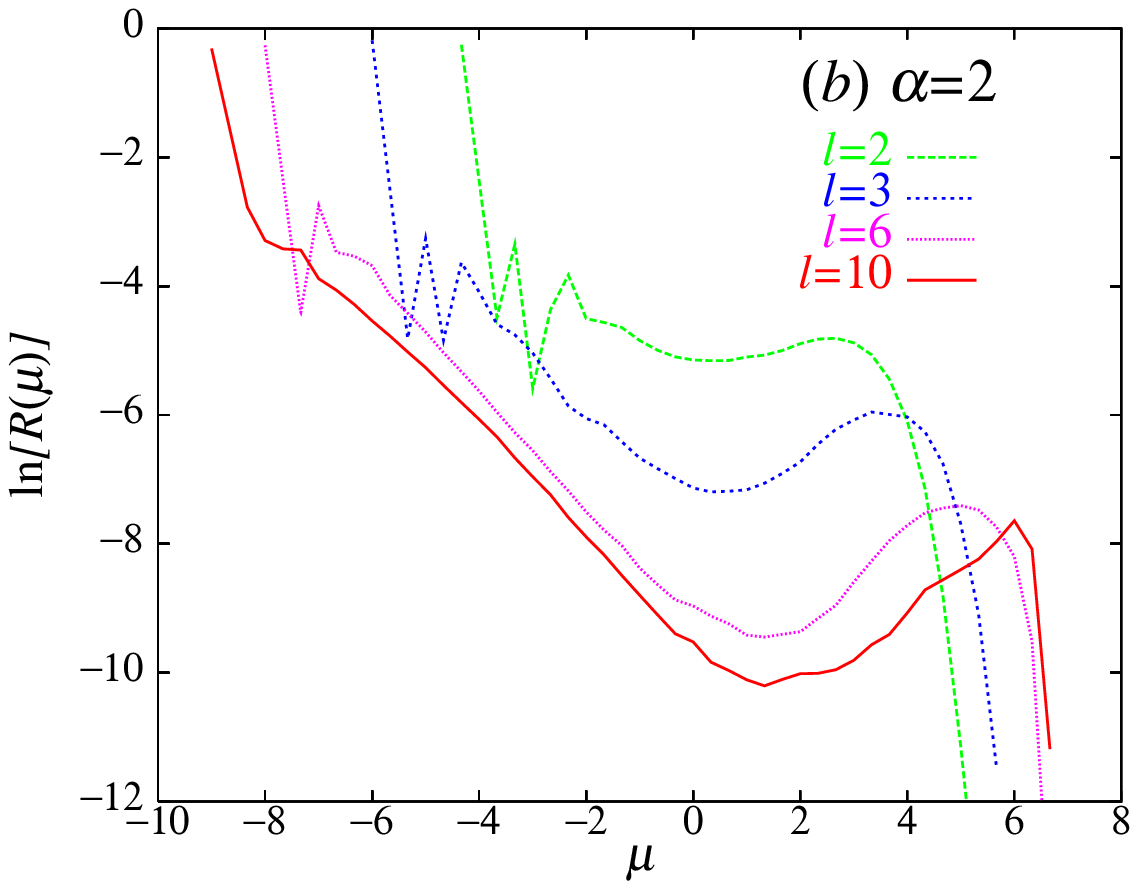}
\end{center}
\caption{
Magnitude distribution of earthquake events for various values of $l$; (a)
for $\alpha =1$ and (b) for $\alpha $=2. The system size is $N=800$.
}
\end{figure}

In Fig.2, we show the magnitude
distribution $R(\mu)$ with
varying the values of the stiffness parameter $l$ as $l=2,3,6$ and 10
for the
case of $\alpha=1$ (Fig.2(a)) and  $\alpha=2$ (Fig.2(b)).
In either case, the form of the magnitude distribution $R(\mu)$ does not
change qualitatively with $l$ in contrast to the case of varying $\alpha$, 
although larger $l$ tends to make the largest possible event even larger. 
This tendency is intuitively easy to understand, since
as the system becomes stiffer the stress release would be made via a
huge event involving larger number of blocks. In the case of $\alpha=1$,
the power-law feature persists for any $l$, the associated
$B$-value being $B\simeq 0.56$ ($l=2$), $B\simeq 0.45$ ($l=6$) and 
$B\simeq 0.45$ ($l=10$). 

We have observed that for $l> 3$ the largest event
sometimes exceeds the system size $N=800$ so that the finite-size effect
on $R(\mu)$ becomes appreciable at larger magnitudes. In order to
control this finite-size effect for larger $l$, we show
in Fig.3 the magnitude
distribution $R(\mu)$ for our largest value of $l$, {\it i.e.\/}, $l=10$ 
for increased lattice sizes of $N=800$, 1,600, 3,200 and 6,400 for the
case of $\alpha=1$ (Fig.3(a)) and of $\alpha=2$ (Fig.3(b)). As can clearly be
seen from the figures, an asymptotic bulk behavior has been reached in our
largest lattice size $N=6,400$, while for smaller $N$ the finite-size effect 
modifies the form of the tail part of the distribution somewhat, though 
the asymptotic
infinite-size behavior obtained in this way does not affect the qualitative 
features of the
magnitude distribution mentioned above.

\begin{figure}[ht]
\begin{center}
\includegraphics[scale=0.65]{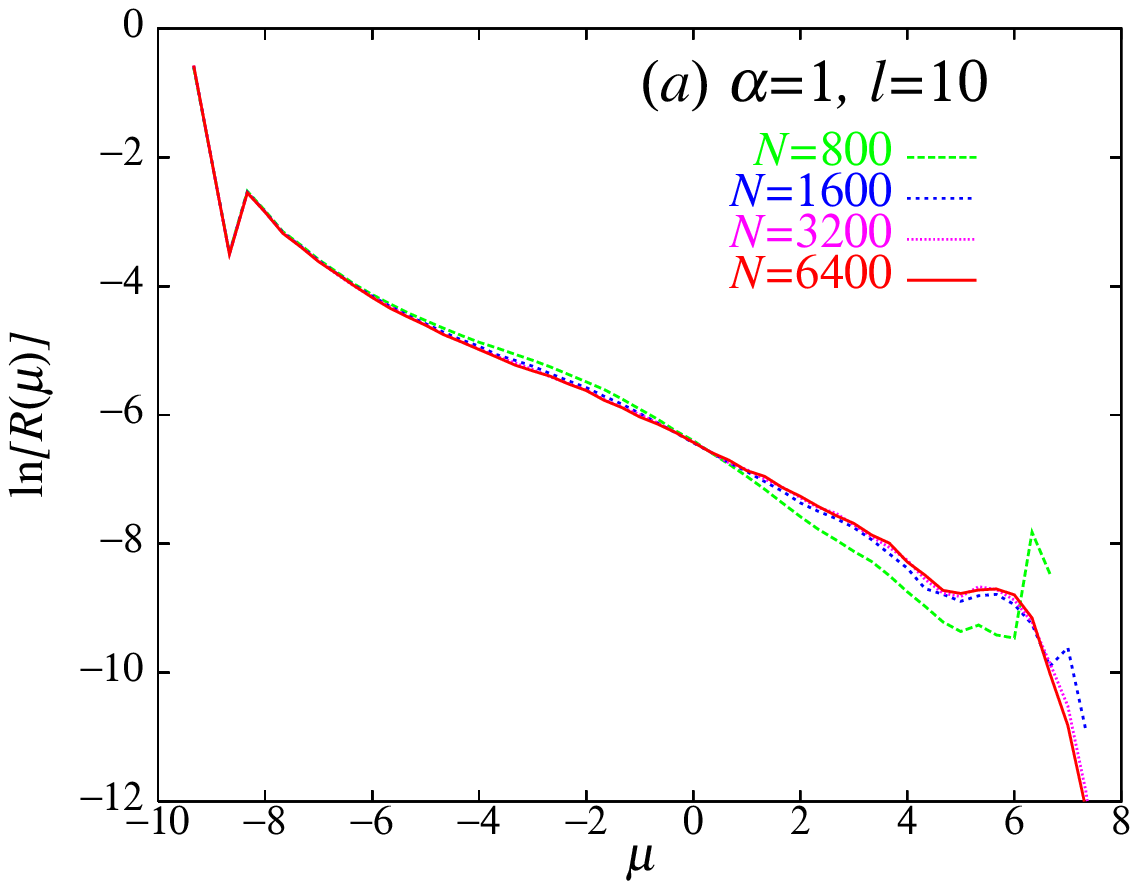}
\includegraphics[scale=0.65]{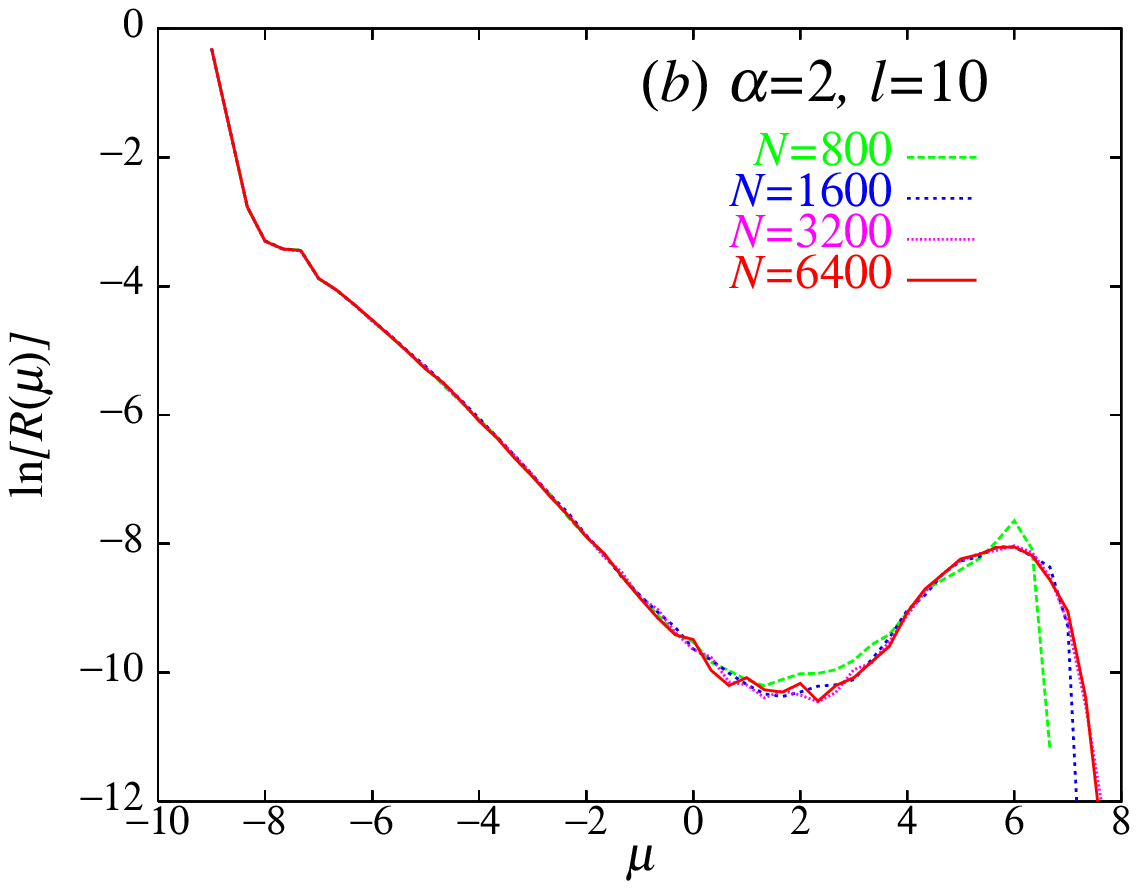}
\end{center}
\caption{
Size dependence of the magnitude distribution of earthquake events 
for the cases
of (a) $\alpha =1$, $l=10$, and of (b) $\alpha $=2, $l=10$. 
}
\end{figure}

\subsection{Mean displacement of blocks in large events}

In order to clarify how the seismic rupture proceeds in each large event, 
we show in Fig.4 the mean displacement of a block which is apart 
from the epicenter of the event 
by distance $r$. The average is made over all events with their magnitudes in the range $4\leq \mu \leq 5$. In the figure, we always take the $r>0$ side to be the side in which the rupture propagates farther, and the $r<0$ side to be the side in which the rupture propagates less, $r=0$ being the epicenter of the event. From Fig.4 one can see the following features: (i) The block
displacement is not maximum at the epicenter, whereas the maximum displacement
occurs somewhat apart from the epicenter. The block which triggers the 
large event does not move much itself. This tendency becomes more pronounced
 for smaller $\alpha$. The ratio of the mean displacement at the epicenter (the block at $r=0$) to the maximum mean displacement is 0.054, 0.70 and 0.74 for $\alpha=1,$ 2 and 3, respectively. (ii) The asymmetry of the curves in Fig.4 
with respect to $r=0$ 
becomes more pronounced for smaller $\alpha$. In the case of $\alpha=1$, 
in particular, the rupture propagates almost in one direction only, and the event occurs as a ``unilateral earthquake''. In other words,
for $\alpha =1$, the epicenter of large events tend to be located
near the edge of the rupture zone. 
This can be confirmed more
quantitatively by calculating the ``eccentricity'' 
of the epicenter $\epsilon =R^*/\bar R$, which is defined
by the ratio of the mean distance between the epicenter and the center of mass
of the rupture zone, $R^*$,
to the mean radius of the rupture zone, $\bar R$. 
Then, we have found $\epsilon = 0.88$, 0.52 and 0.53
for the cases $\alpha =1$, 2 and 3, respectively.
The occurrence of unilateral earthquakes for smaller $\alpha$
may naturally be understood if one notes that the relative
weakness of the stick-slip instability prevents the initiated rupture
from propagating far into both directions.

\begin{figure}[ht]
\begin{center}
\includegraphics[scale=0.65]{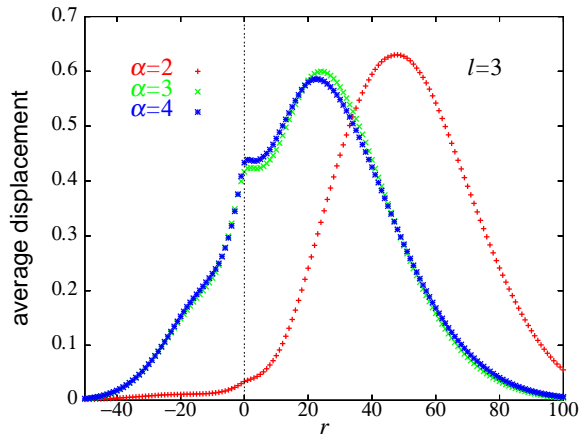}
\end{center}
\caption{
Mean displacement of blocks in large events with their magnitudes in the range $4 \leq \mu \leq 5$ plotted versus the distance of a block 
from the epicenter of the event. The $r>0$ side is taken to be the side in which the rupture propagates farther, and the $r<0$ side to be the side in which the rupture propagates less, $r=0$ being the epicenter of the event.  The parameter $\alpha$ is $\alpha=1,2$ and 3. The system size is $N=800$.
}
\end{figure}

\subsection{Local recurrence-time distribution}

A question of general interest 
may be how large earthquakes repeat in time, 
do they occur near periodically or irregularly? 
One may ask this question either locally, {\it i.e.\/}, for a given finite 
area on the fault,
or globally,  {\it i.e.\/}, for an entire fault system.
The picture of characteristic
earthquake presumes the existence of a characteristic recurrence time. 
In that case, the distribution of the recurrence time of large earthquakes, 
$T$, 
is expected to exhibit a peak structure at such a characteristic time scale.
If the SOC concept applies to large earthquakes, on the other hand, 
such a peak structure would not show up. 

In Fig.5(a), we show the distribution of the
recurrence time $T$ of large earthquakes with $\mu \geq \mu _c=3$, 
measured locally for the case of $l=3$.
In the insets, the same data including the tail part 
are re-plotted on a semi-logarithmic scale.
In defining the recurrence time locally, 
the subsequent large event is counted when a large
event occurs with its epicenter in the region within
30 blocks from the epicenter of the previous large event.
The mean recurrence time $\bar T$ is then estimated to be
$\bar T\nu =1.47$, 1.12, and 1.13 for $\alpha=1$, 2 and 3,  respectively.

The local recurrence-time
distribution shown in Fig.5(a) has the following two noticeable features. 
(i) The tail of the distribution  is exponential at longer $T> \bar T$
for all values of $\alpha$. 
Such an exponential tail of the distribution has also been
reported for real faults [\textit{Corral}, 2004].
(ii) The form of the distribution at shorter $T< \bar T$ 
is non-exponential,
and largely differs between
for $\alpha =1$ and for $\alpha =2$ and 3. 
For $\alpha=2$ and 3, the distribution
has an eminent  peak at around
$\bar T\nu \simeq 0.5$, not far from the mean  
recurrence time.
This means the existence of a characteristic recurrence time,
suggesting the near-periodic recurrence of 
large events. Indeed, such a near-periodic recurrence of large
events was reported for several real faults [\textit{Nishenko and
Buland}, 1987;\textit{Scholz}, 1990].

\begin{figure}[ht]
\begin{center}
\includegraphics[scale=0.65]{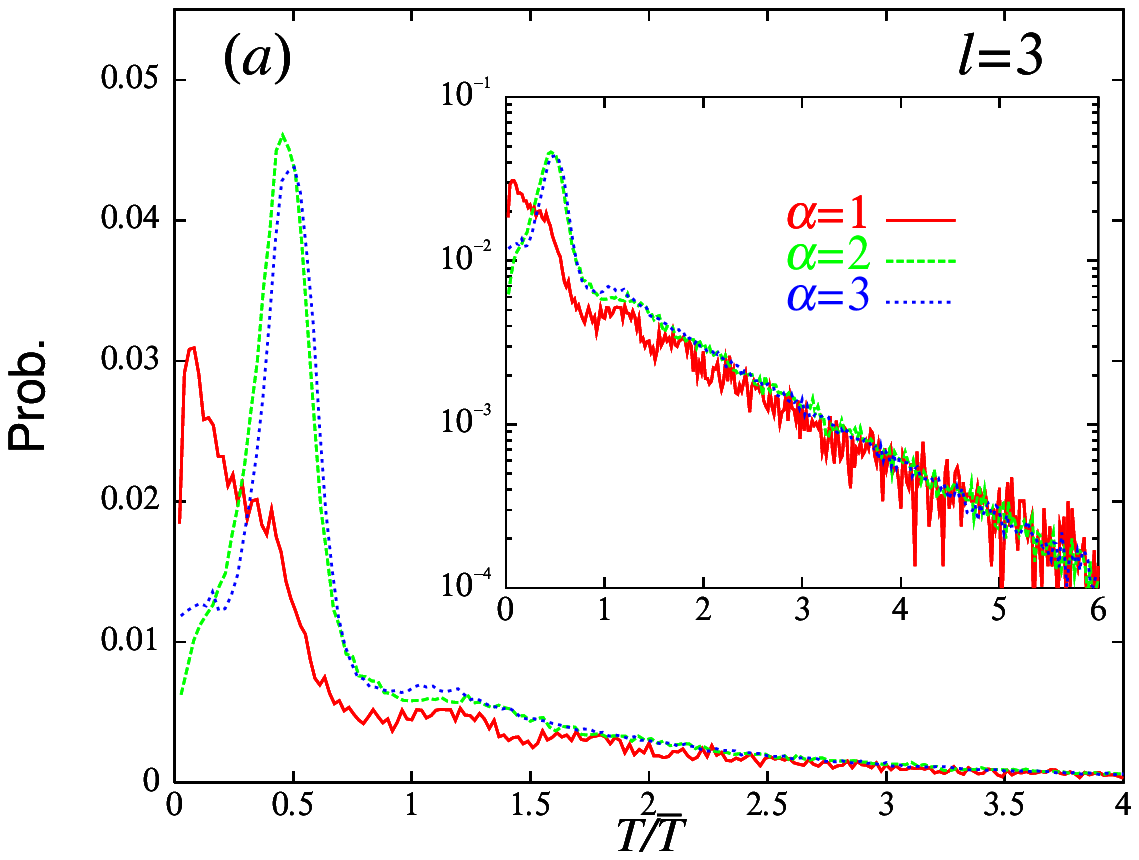}
\includegraphics[scale=0.65]{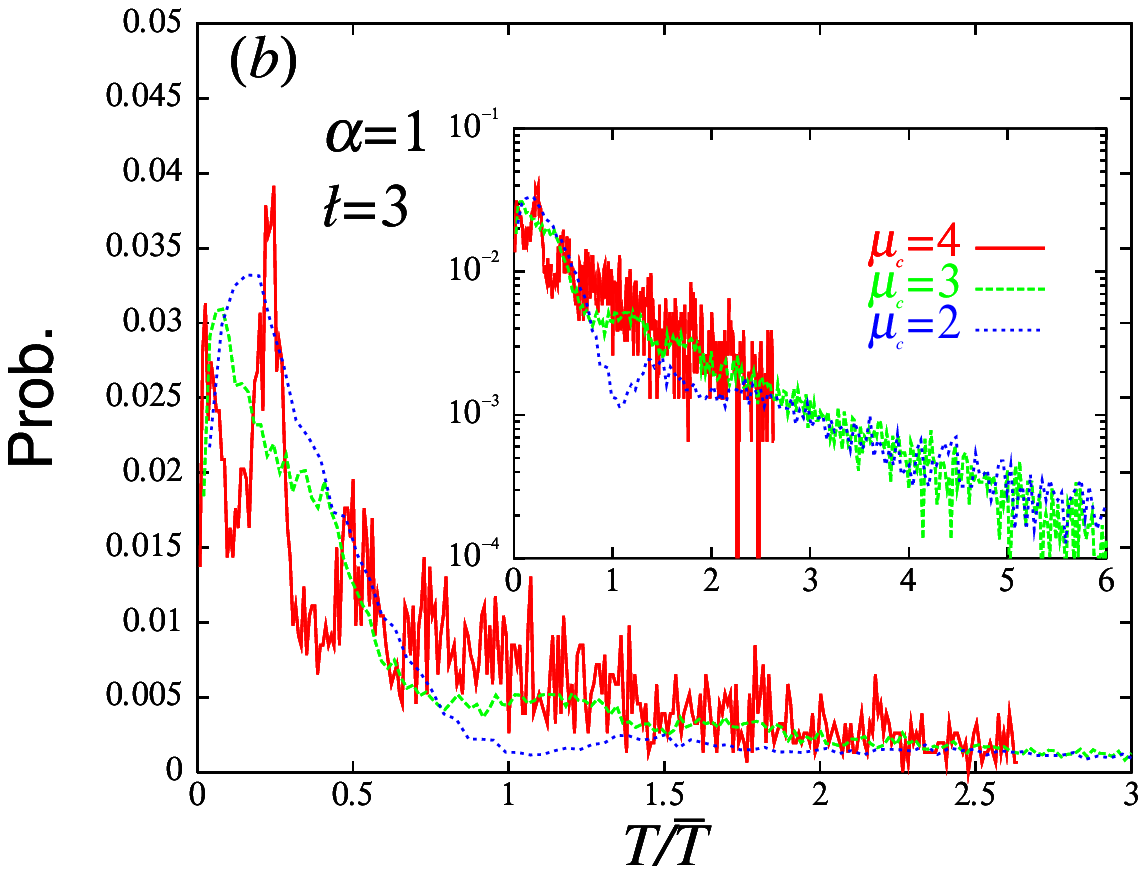}
\end{center}
\caption{
The local recurrence-time distribution of large events. The recurrence 
time $T$ is normalized by its mean $\bar T$. In (a), the parameter
$\alpha$ is varied with fixing $l=3$ with the magnitude threshold $\mu_c=3$.
The mean recurrence times are $\bar T\nu =1.47$, 1.12, and 1.13 
for $\alpha=1$, 2 and 3,  respectively.
In (b), the magnitude threshold $\mu_c$ is varied with fixing $\alpha=1$ and 
$l=3$. The mean recurrence times are 
$\bar T\nu =0.76$, 1.47, and 3.42 for $\mu_c=2$, 3 and 4, respectively. 
In each figure,
the insets represent the semi-logarithmic plots including the tail part of 
the distribution. The system size is $N=800$.
}
\end{figure}

For $\alpha =1$, by contrast,
the peak located close to the mean $\bar T$ is hardly discernible. 
Instead, the distribution has a pronounced peak at a 
shorter time $\bar T\nu \simeq 0.10$, just after the previous large event.
In other words, large events for $\alpha=1$ tend to occur as 
 ``twins''. This has also been confirmed by our analysis of the time record
of large events.

As was noticed in Fig.4, a large event for the case of $\alpha=1$
often occurs as a ``unilateral earthquake'' where the rupture propagates
only in one direction, hardly propagating 
in the other direction.
When a large earthquake occurs in the form of  
such a unilateral earthquake, further
loading due to the plate motion tends to trigger the subsequent
large event in the opposite direction, causing a twin-like event. 
This naturally explains the small-$T$ peak observed in Fig.5(a) for $\alpha=1$.

The distribution functions for $\alpha =2$ and 3 can be
well fitted by the sum of the exponential representing the tail part and the Gaussian representing the peak part, 
\begin{equation}
\approx C_1\exp [-\frac{1}{2}(\frac{T-T_1}{\Delta T})^2] + C_2\exp [- \frac{T}{T_2}],
\end{equation}
where $T_1\nu\simeq 0.51$, $\Delta T\nu \simeq 0.12$, $T_2\nu \simeq 1.40$ for $\alpha=2$, and $T_1\nu \simeq 0.57$, $\Delta T\nu \simeq 0.12$, $T_2 \nu \simeq 1.45$ for $\alpha=3$, respectively.
For $\alpha =1$, by contrast, the short-time part cannot be
fitted by the Gaussian, although 
the tail part of the distribution can be well fitted by
the exponential with $T_2 \nu \simeq 1.98$. 

In Fig.5(b), the
recurrence-time distribution of large events is shown for the case of 
$\alpha =1$ and $l=3$, with varying the 
magnitude threshold as $\mu _c=2$, 3 and 4. As can be seen form Fig.5(b), 
the form of the distribution for
$\alpha=1$ largely changes with the threshold value $\mu_c$.
Interestingly, in the case of  $\mu_c=4$, 
the distribution has {\it two\/} 
distinct peaks, one corresponding to 
the twin-like event and the other to the near-periodic event.
Thus, even in the case of $\alpha=1$ where the critical features
are apparently dominant for smaller thresholds $\mu_c=2$ and 3, features of 
a characteristic earthquake becomes 
increasingly eminent when 
one looks at very large events.

By contrast,  the form of the recurrence-time
distribution for $\alpha=2$ turns out to be 
rather insensitive to the threshold value 
$\mu_c$, always keeping a pronounced peak structure (the figure not shown). 
The peak position varies
with $\mu_c$ considerably, however, reflecting the change of the mean 
recurrence time $\bar T$ with $\mu_c$, {\it i.e.\/}, $\bar T\nu \simeq 0.76\ (\mu_c=2)$, $\bar T\nu \simeq 1.12\ (\mu_c=3)$ and $\bar T\nu \simeq 2.58\ 
(\mu_c=4)$.

\begin{figure}[ht]
\begin{center}
\includegraphics[scale=0.65]{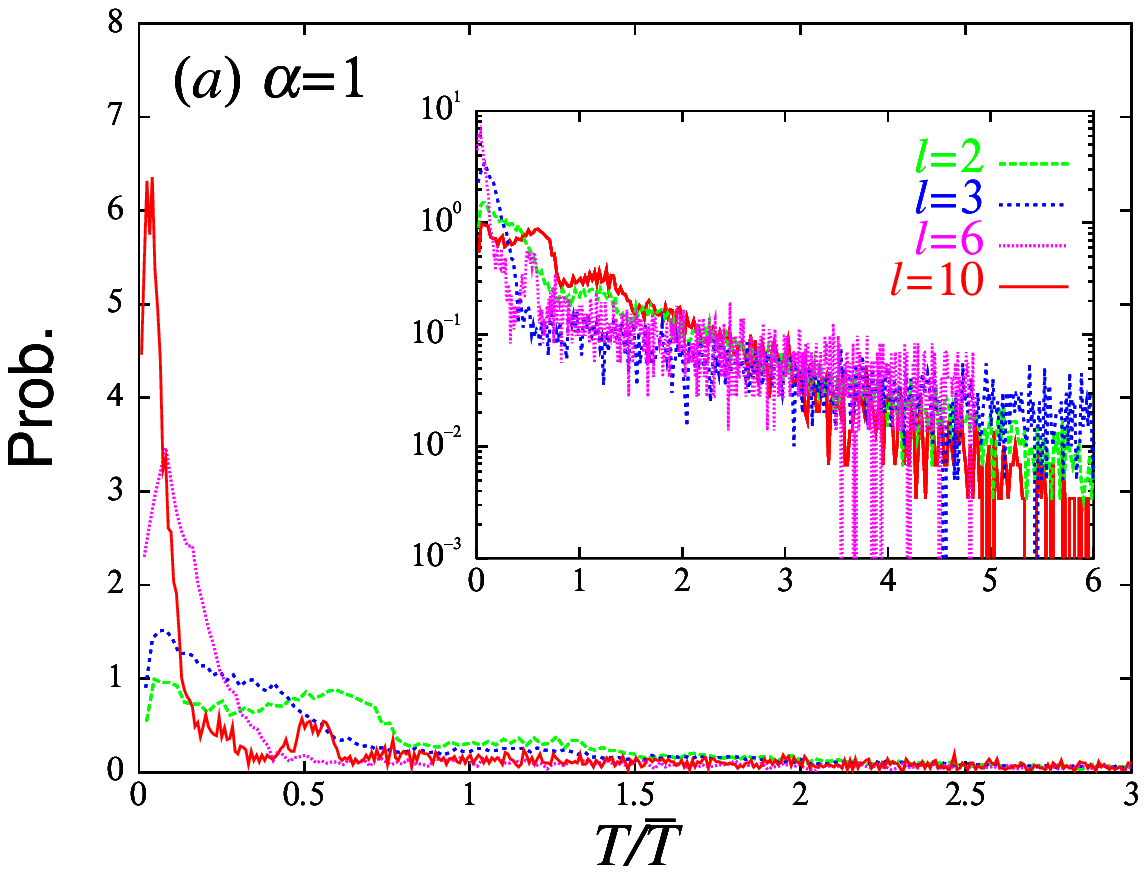}
\includegraphics[scale=0.65]{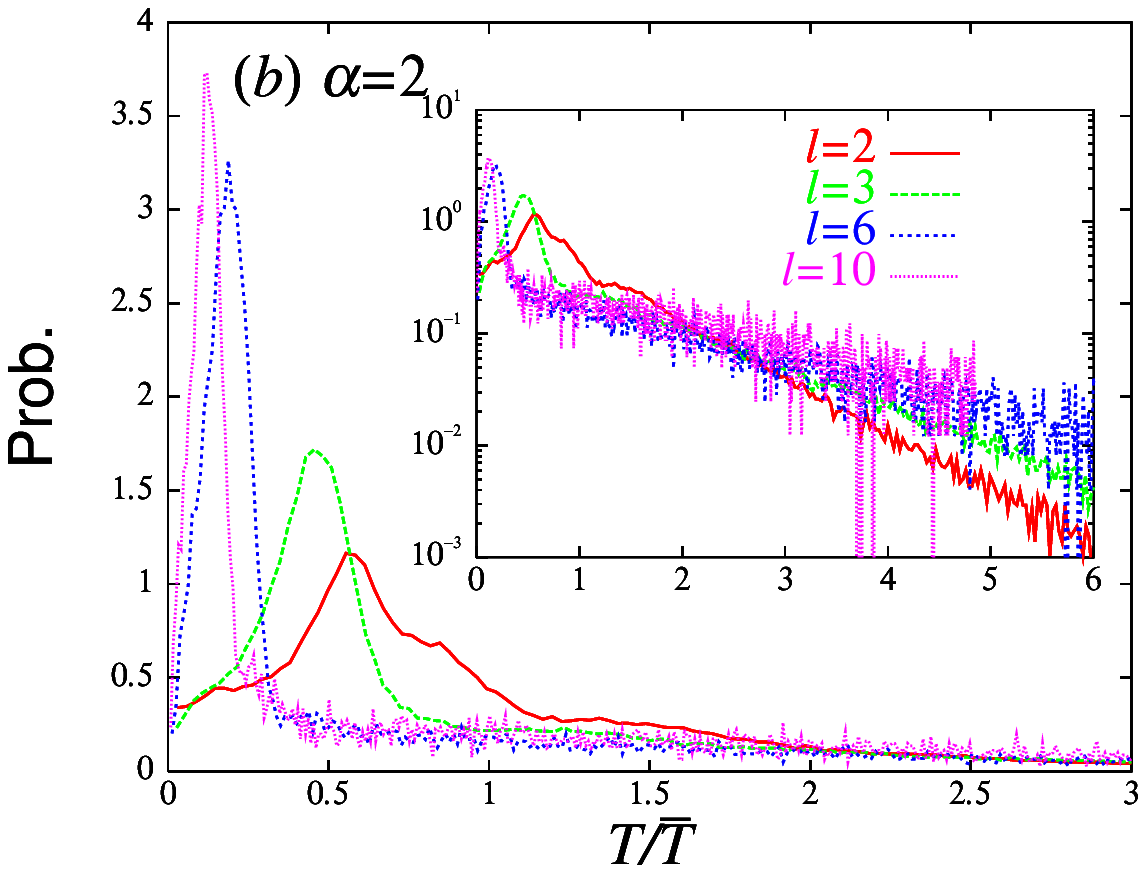}
\end{center}
\caption{
The local recurrence-time distribution of large events with their magnitudes
greater than $\mu_c=3$
for various values of $l$, for the cases of $\alpha=1$ (a) and of $\alpha=2$
(b).
The recurrence time $T$ is normalized by its mean $\bar T$, which are
 $\bar T\nu =1.30$, 1.47, and 1.64 for $l=2$, 3 and 4 
in the case of $\alpha=1$, while 
$\bar T\nu =1.03$, 1.12, and 1.47 for $l=2$, 3 and 4 in the case of
 $\alpha=2$.
The insets represent the semi-logarithmic plots including the tail part of the distribution. The system size is $N=800$.
}
\end{figure}

In Fig.6, we show
the local recurrence-time distribution of large events (with 
the threshold value $\mu_c=3$ here) for various values of 
the parameter $l$, for the cases of
$\alpha=1$ (a) and  of $\alpha=2$ (b).
As can be seen from the figures, 
qualitative features of the recurrence-time distribution are not very 
sensitive to the $l$-value, although periodic features tend to be
enhanced for larger $l$.

\subsection{Global recurrence-time distribution}

In Fig.7(a), we show
the {\it global\/} recurrence-time distribution of large events
with $\mu_c=3$,
{\it i.e.\/}, the one for an entire
fault system with $N=800$ for various values of $\alpha $.
The parameter $l$ is fixed to be $l=3$.  For this rather small value of $l$, 
the system size $N=800$ is sufficiently large.
As can clearly be seen from the figure, the form of the
distribution takes a 
different form from the local one.
The peak structure seen in the local distribution no longer 
exists here. Furthermore, 
the form of the distribution tail at larger $T$
is no longer a simple exponential, faster than exponential:
See a curvature of the data in the inset of Fig.7(a). 

In Fig.7(b), we show the $l$-dependence of the global recurrence-time 
distribution of large events with $\mu_c=3$ for the case of $\alpha=2$
for a fixed system size $N=800$.
As can clearly be seen from the figure, the form of the distribution
changes dramatically for larger $l$. In the case of $l=10$, in particular,
the distribution exhibits a pronounced 
peak structure suggesting a near-periodic
occurrence of large events. Similar peak structure of the global 
recurrence-time distribution was also reported in [\textit{Carlson}, 1991a] 
for the case of $\alpha=2.5$ and 
$l=10$ with $N=100$. Care has to be taken to the possible 
finite-size effect here. In Fig.7(c), we show the size dependence 
of the global recurrence-time distribution of large events with $\mu_c=3$ 
for the case of $\alpha=2$ and $l=10$ with varying the system size
in the range  $800\leq N \leq 6,400$. 
The data for the smallest size $N=800$
corresponds to the one shown in Fig.7(b) for $l=10$. As can clearly be seen
from Fig.7(c), as the system size $N$ is increased, the peak structure in 
the distribution goes
away, eventually 
giving way to a monotonic distribution quite similar to the ones
observed for $l=3$ in Figs.7(a) and (b).
This observation indicates that the monotonic behavior as observed in Fig.7(a)
is a general feature of the global recurrence-time distribution of the
1D BK model so long as the system size is taken sufficiently large.

The present observation that the local and the global recurrence-time
distributions of the 1D BK model 
exhibit mutually different behaviors in bulk system
means that the form of the
distribution  depends on the length scale
of measurements, as well as on the $\alpha$-value. In order to see
such a behavior for lager $l$, however, 
care has to be taken so that the system size is large
enough.
Such scale-dependent features of the
recurrence-time distribution of the BK model is in apparent contrast with the
scale-invariant features of the
recurrence-time distributions recently reported for some of real faults 
[\textit{Bak et al.}, 2002;\textit{Corral}, 2004].

\begin{figure}[ht]
\begin{center}
\includegraphics[scale=0.65]{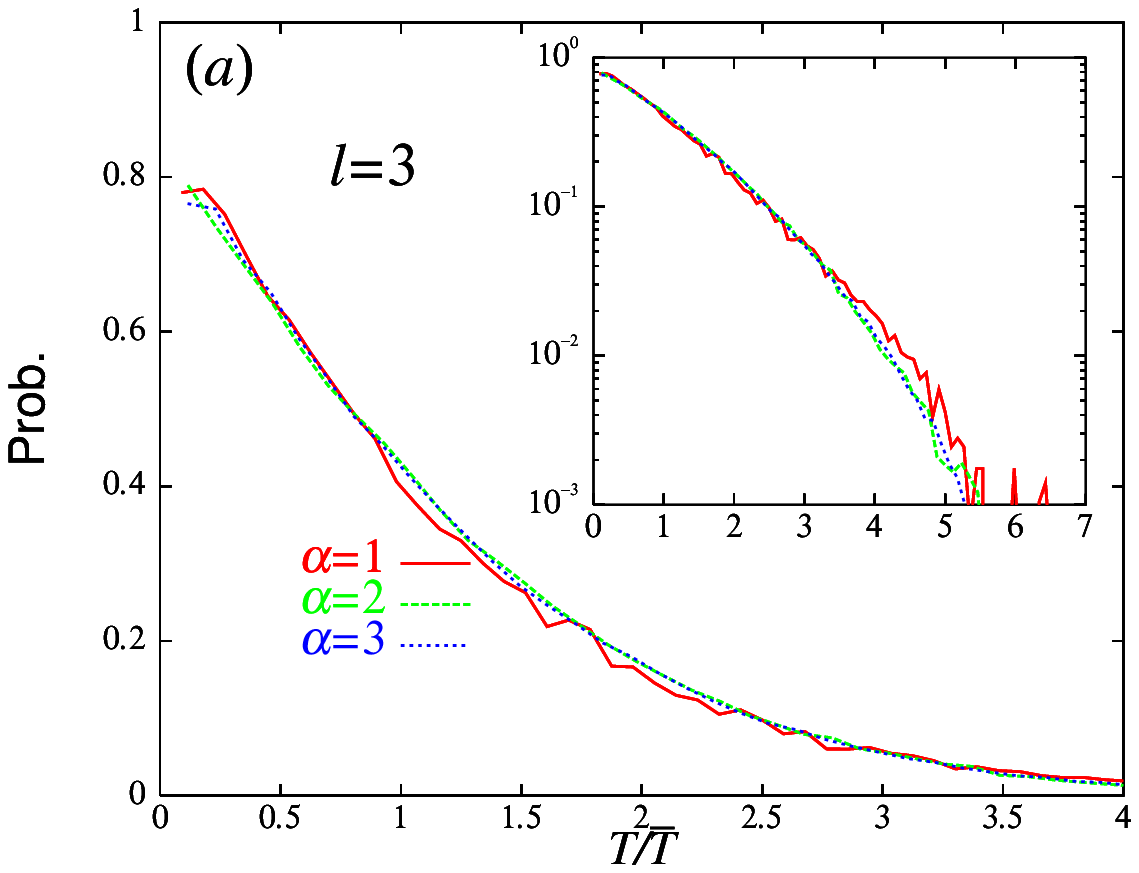}
\includegraphics[scale=0.65]{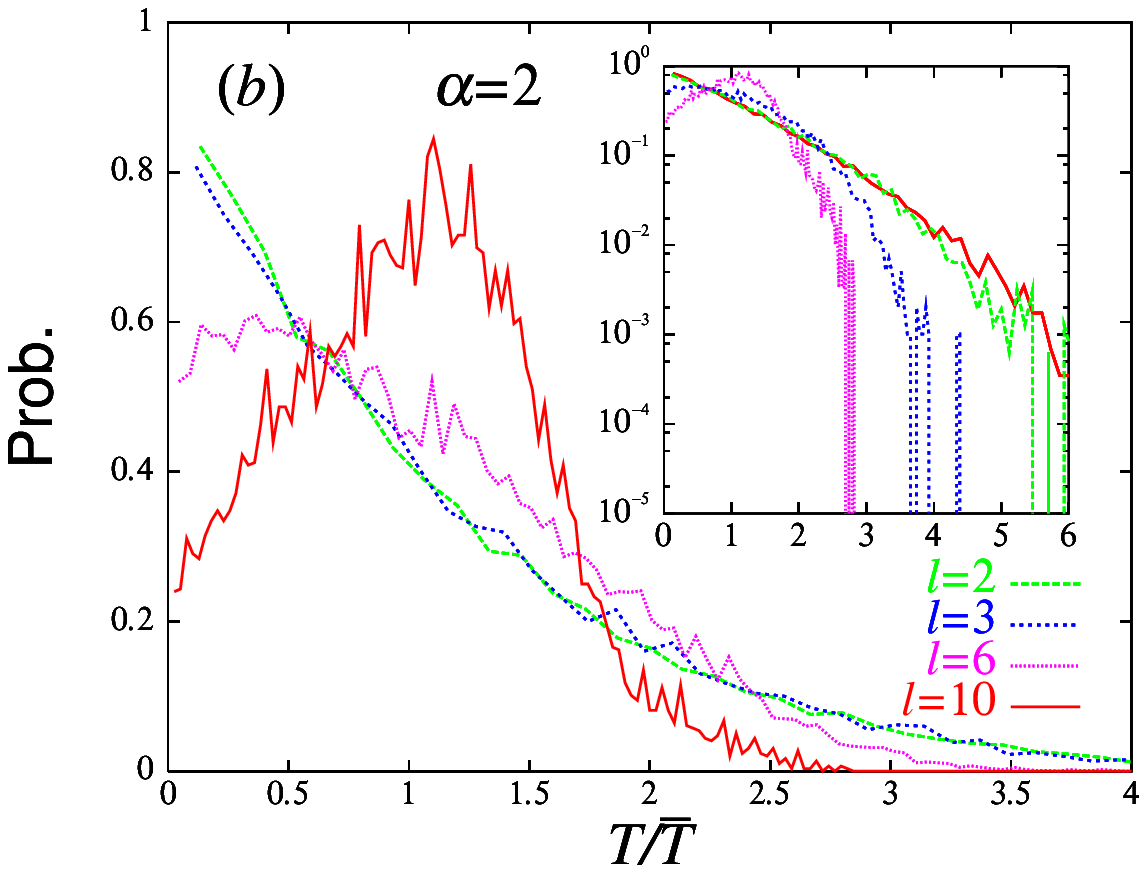}
\includegraphics[scale=0.65]{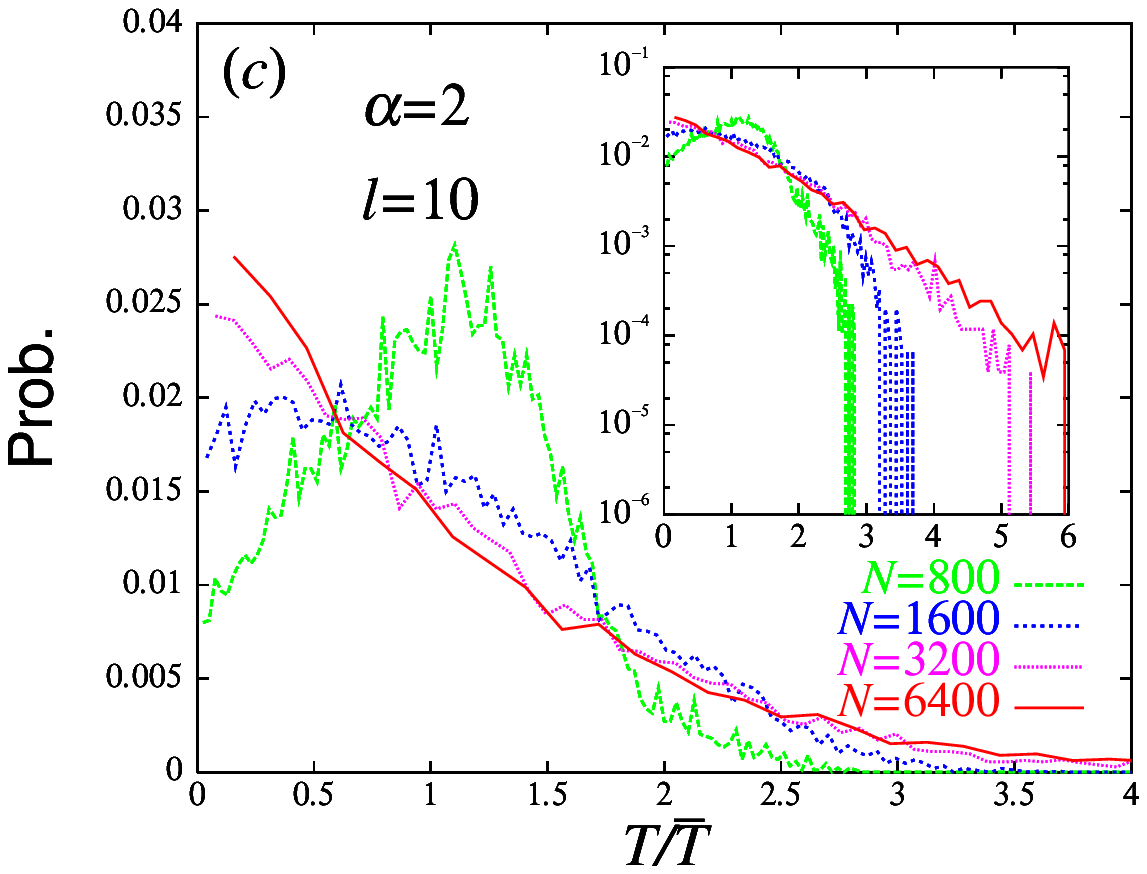}
\end{center}
\caption{
The global recurrence-time distribution of large events with their magnitudes
greater than $\mu_c=3$.
The recurrence time $T$ is normalized by its mean 
$\bar T$. In (a),
the parameter $\alpha$ is varied with fixing $l=3$ for the size $N=800$. 
The mean recurrence times  are
$\bar T\nu =0.011$, 0.086 and 0.087 for $\alpha=1$, 2 and 3,  respectively.  
In (b), the parameter $l$ is varied with fixing $\alpha =2$ 
for the size $N=800$. Pronounced peak structure is observed for larger $l$
for this size.
In (c), the size $N$ is varied for the case of $\alpha=2$ and $l=10$.
A significant finite-size effect is evident for smaller $N$, whereas, for 
sufficiently large
$N$, the global recurrence-time distribution becomes monotonic which resembles 
the one observed in (a) for the case of smaller $l$.
In each figure, 
the insets represent the semi-logarithmic plots including the tail part of 
the distribution.
}
\end{figure}

\subsection{Magnitude correlations between successive large events}

In the previous subsections, we studied the recurrence time,
{\it i.e.\/}, the time elapsed between 
two successive large events. In this subsection, we study the
correlation between the magnitudes of the two successive large events.
There are two common views in the public
concerning the magnitudes of the two successive
large events. One is that there is no correlation between 
the magnitudes of the two successive large events. This view
underlies, {\it e.g.\/}, 
the so-called time-predictable and size-predictable models of earthquakes 
[\textit{Shimazaki and Nakata},1980; \textit{Scholz}, 1990].
The other is that there is an {\it anti-correlation\/} between 
the magnitudes of the two successive large events, {\it i.e.\/}, when the first event happens to be not so large, then the next
event tends to be larger, and vice versa. 
Naively, this tendency is a natural expectation since, when the first event 
is only moderate, the stress accumulated in the region might not fully be released making the next event even larger.

\begin{figure}[ht]
\begin{center}
\includegraphics[scale=0.65]{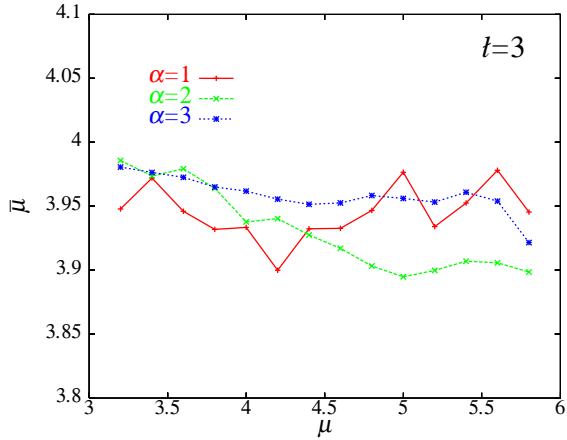}
\end{center}
\caption{
The mean magnitude of large events with their magnitudes
greater than $\mu_c=3$ plotted versus the  magnitudes of the last large
event preceding this event. The parameter $\alpha$ is either
$\alpha=1$, 2 or 3, while $l$ is fixed to be $l=3$. The system size is $N=800$.
}
\end{figure}

In Fig.8, we show  for each case 
of $\alpha =1$, 2 and 3 the mean magnitude of large events with
$\mu_c=3$ versus the  magnitude values of the large
event preceding this event.  As in  Figs.5-7, the successive large event
is identified when an event of its magnitude $\mu > \mu_c=3$ occurs with its epicenter lying in the region
less than 30 blocks from the epicenter of the preceding large event. 
The parameter $l$ is
taken to be $l=3$ here, while essential features remain the same for other 
values of $l$.

As can be seen from Fig.8,  correlation between the magnitudes of 
the successive large events  is hardly
appreciable in the case of $\alpha =1$, while, 
in the cases of $\alpha =2$ and 3, 
there is a weak anti-correlation between the magnitudes of
the two successive events, {\it i.e.\/}, the next event tends to be larger when the first event is moderate, and vice versa. 
Hence, there seems to be a weak 
tendency that the larger $\alpha$-value induces 
an anti-correlation between the magnitudes of the two
successive large events.

In Fig.9, we show the mean magnitude 
of large events versus the magnitude of the large
event preceding this event for several values of $l$,
with fixing $\alpha=2$.
As can be seen from the figure, there is a weak
tendency that the larger $l$-value
induces an anti-correlation between the magnitudes of the two
successive large events. 

If one combines the observation here with those of 
the previous subsections, one may conclude that there is no correlation between the  magnitudes of the two successive events 
when an earthquake is ``critical'' (as it should be), whereas there appears an anti-correlation  between the  magnitudes of the two successive events when an earthquake is ``characteristic''.

\begin{figure}[ht]
\begin{center}
\includegraphics[scale=0.65]{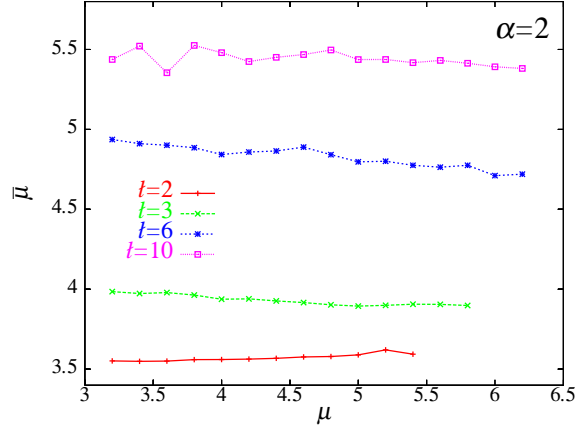}
\end{center}
\caption{
The mean magnitude of large events with their magnitudes
greater than $\mu_c=3$ plotted versus the  magnitudes of the last large
event preceding this event. The parameter $l$ is either $l=2$, 3, 6 or 10, where $\alpha$ is fixed to be $\alpha=2$. In the case of smaller $l$,
there is virtually no event with very large magnitude.
The system size is $N=800$.
}
\end{figure}

\subsection{Time correlation of events associated with the mainshock}

In Fig.10(a), we show the time correlation function
between large events and
events of arbitrary sizes (dominated in number by small events)
for various values of the parameter $\alpha$ with fixing $l=3$. More precisely,
we plot the mean number of events of arbitrary sizes occurring within 30 blocks from the epicenter of the
mainshock before  ($t<0$) and after
($t>0$) the mainshock, where the occurrence of the mainshock is taken to be 
the origin of the time $t=0$. The average is taken over all large events of their magnitudes with $\mu \geq \mu_c=3$.
Note that, by this way of event counting, when
more than one large event occurs successively close in time, the 
same event might be counted more than once in the time correlation function 
at distinct times $t$, 
each associated with distinct large events.  
The number of events are counted here with the
time bin of $\Delta t\nu =0.02$.

As can be seen from Fig.10(a), after the mainshock, 
the seismic activity stays at lower level for some period,
which corresponds to the calm period or quiescence. 
In the cases of $\alpha=2$ and 3, this calm period continues during about half of the recurrence time
of large events.
After this calm period,
the seismicity is gradually activated, 
eventually leading to the next large event, which manifests itself 
in the peak of the time correlation function at around $t\simeq \bar T$. Thus,
for the cases of $\alpha=2$ and 3,
the interval period between the two large events can be divided into two
halves: The former half is the calm period where the seismic activity stays 
at lower level, while 
the latter half is the active period where the seismicity attains higher 
level. 

In the case of $\alpha=1$, by contrast, the calm period after the 
mainshock is very short, and the active seismicity resumes soon after the
mainshock. This immediate seismic activation after the mainshock  in the case of $\alpha=1$ may be understood as associated with the 
next large event which tends to occur immediately after the large event as
a ``twin''-like event: See Fig.5(a).

\begin{figure}[ht]
\begin{center}
\includegraphics[scale=0.65]{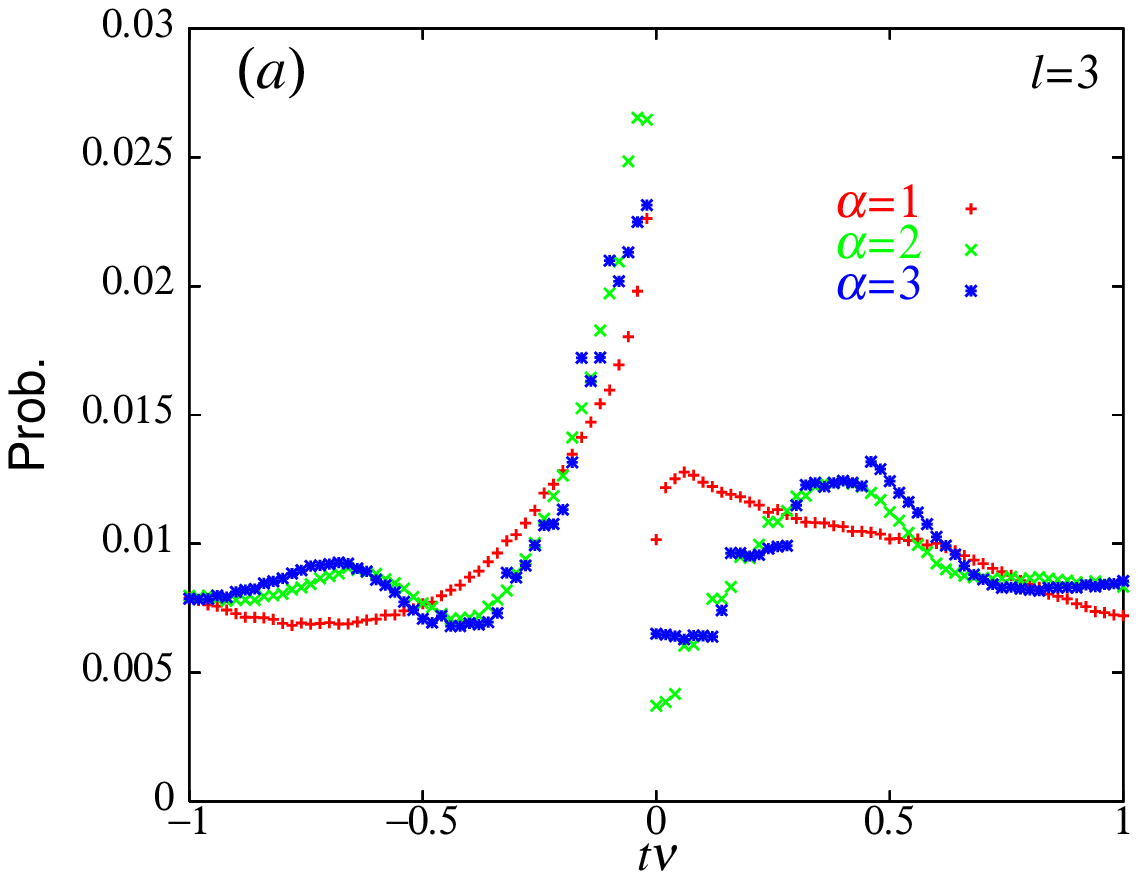}
\includegraphics[scale=0.65]{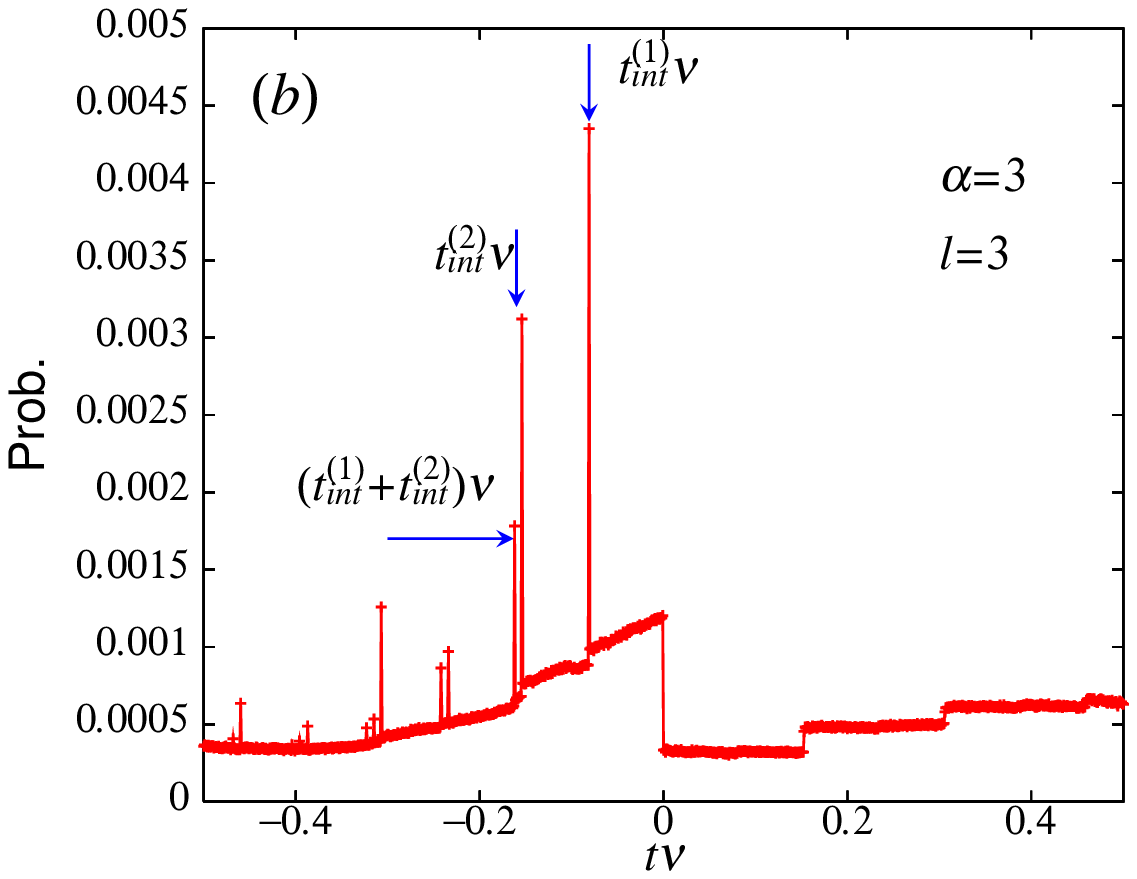}
\end{center}
\caption{
The time correlation function 
between large events occurring at time $t=0$ and
events of arbitrary sizes (dominated in number by small events) occurring at
time $t$
for various values of $\alpha$. 
Events of arbitrary sizes occurring within 30 blocks from the epicenter of the
large event are counted. 
The negative time $t<0$ represents the time before the mainshock, while the positive time $t>0$ represents the time after the mainshock. The average is taken over all large events with its magnitude $\mu >\mu_c=3$. In (a), 
the number of events are counted with the
time bin of $\Delta t\nu =0.02$, while in (b), it is counted with the higher
time resolution of $\Delta t\nu =0.001$. The system size is $N=800$.
}
\end{figure}

Before the mainshock ($t<0$),
a remarkable acceleration of seismic activity 
occurs irrespective of the $\alpha$-value.
Hence, the existence of the active period preceding the mainshock
appears to be 
a general property of the BK model.

Closer inspection of Fig.10(a) reveals that the calculated
time-correlation function exhibits irregular structures, particularly in
the case of $\alpha=3$. Thus,
we show in Fig.10(b) the time-correlation function for $\alpha=3$ 
{\it with higher time resolution\/} 
than in Fig.10(a), {\it i.e.\/}, the number of events are counted here with the smaller time bin of $\Delta t\nu =0.001$. Now, it can 
clearly be seen from
Fig.10(b) that several ``spikes'' appear in the time correlation function. 
It turns out that
all these spikes consist almost exclusive of one-block events.
Furthermore,  all the interval
times between these spike-like events and the mainshock at $t=0$ are
given by the linear combination of the two elementary
interval times $t_{int}^{(1)}$ and
$t_{int}^{(2)}$ 
as $n_1t_{int}^{(1)} + n_2t_{int}^{(2)}$ ($n_1$ and $n_2$ integers): 
See the arrows in Fig.10(b). 

In the present model, one can see that the block motions in
all possible {\it one-block\/} events
are exactly the same when measured from their stopping sites 
irrespective of the positions of the neighboring
blocks so long as only one block is involved in the event. 
This is an immediate consequence  of the equation of motion of the 
model (2).
Now, suppose that there occurs a one-block event at the block $i$
while  the same block $i$ triggers the next event  some
time after this one-block event, {\it i.e.\/},
the block $i$ happens to be an epicenter of the next event which may be
either one-block event or many-blocks event.
When the two neighboring blocks $i\pm 1$ are kept stopped during this 
time period, the interval time between the one-block event and the next
event triggered by the  block $i$ is uniquely determined for
all possible one-block events, which is 
equal to $t_{int}^{(2)}$ given above. On the other hand, when one
of the neighboring blocks, {\it i.e.\/}, either the one at $i-1$ or at 
$i+1$ exhibits a one-block event during this time period, 
the interval time is again uniquely determined for
all possible one-block events, and is 
equal to $t_{int}^{(1)}$ given above.  
These two interval times 
$t_{int}^{(1)}$ and $t_{int}^{(2)}$, {\it i.e.\/}, the interval times between
the one-block event and the next event of arbitrary sizes triggered 
by the same block, are calculable numerically from the equation
of the motion (2) as a function of the parameters $\alpha$ and $l$.
The results
are shown in Fig.11. Both $t_{int}^{(1)}$ and $t_{int}^{(2)}$ increase with
increasing $\alpha$. In fact, the interval times between the spikes
observed  in Fig.10(b) can be well described by the linear combination of the
calculated $t_{int}^{(1)}$ and $t_{int}^{(2)}$: In the case of $\alpha=3$
and $l=3$ corresponding to Fig.10(b), $t_{int}^{(1)}$ and $t_{int}^{(2)}$ are
calculated to be $t_{int}^{(1)}\nu =0.081$ and $t_{int}^{(2)}\nu =0.154$.

Although the existence of such characteristic interval times associated with
the one-block event, $t_{int}^{(1)}$ and $t_{int}^{(2)}$, is an inevitable
attribute of the present model, it is not clear at the present stage
whether this property of the BK model
has a counterpart in real earthquakes in some way or other. Further studies
are required to clarify this point.

\begin{figure}[ht]
\begin{center}
\includegraphics[scale=0.65]{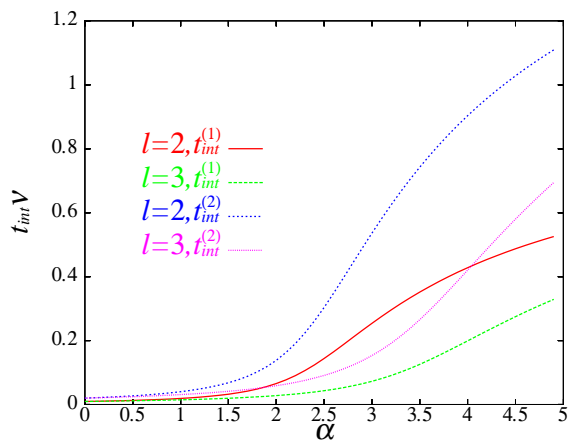}
\end{center}
\caption{
The two elementary interval times 
$t_{int}^{(1)}$ and $t_{int}^{(2)}$, {\it i.e.\/}, the interval times between
the one-block event and the next event of arbitrary sizes triggered 
by the same block,  calculated from the equation
of motion  as a function of the parameters $\alpha$ and $l$. For more
detailed definition, see the text.
}
\end{figure}

\subsection{Spatial correlations of events before the mainshock}

In this subsection, we examine the time-development of 
spatial correlations of events
preceding the mainshock. 
Fig.12 represents the spatial correlation function 
between the mainshock and the preceding 
events of arbitrary sizes (dominated in number
by smaller events) for several time periods  before the mainshock, for 
various cases of the parameters, {\it i.e.\/},
$\alpha=1,\ l=3$ (a), $\alpha=2,\ l=3$ (b),  $\alpha=1,\ l=10$ (c), and
$\alpha=2,\ l=10$ (d): 
It represents the conditional probability that,
provided that a large event with $\mu >\mu_c =3$ 
occurs at a time $t_0$ and at a spatial
point $r_0$,
an event of arbitrary sizes occurs at a time $t_0-t$ and at a spatial 
point $r_0\pm r$. The calculated spatial correlation functions  
for the cases of $\alpha =1,2$ and 3
are shown as a function of $r$.

\begin{figure}[ht]
\begin{center}
\includegraphics[scale=0.65]{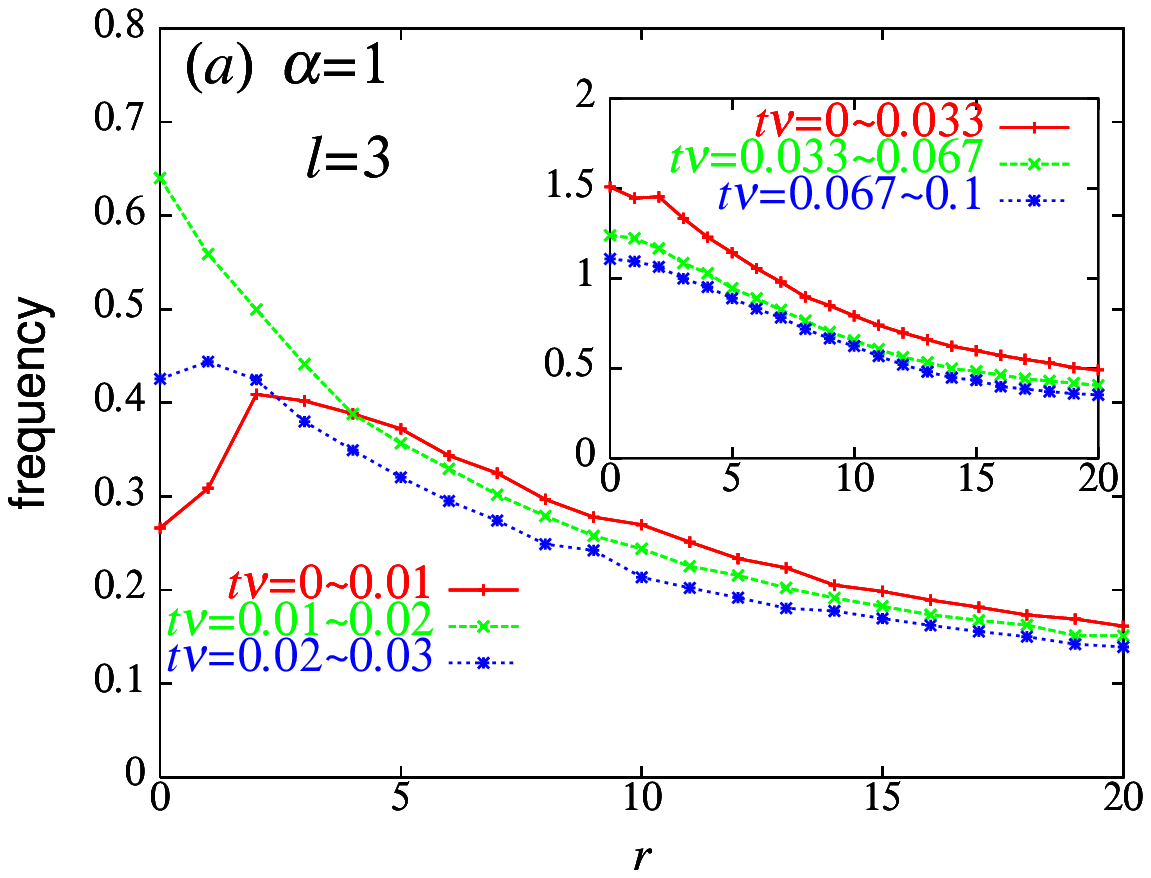}
\includegraphics[scale=0.65]{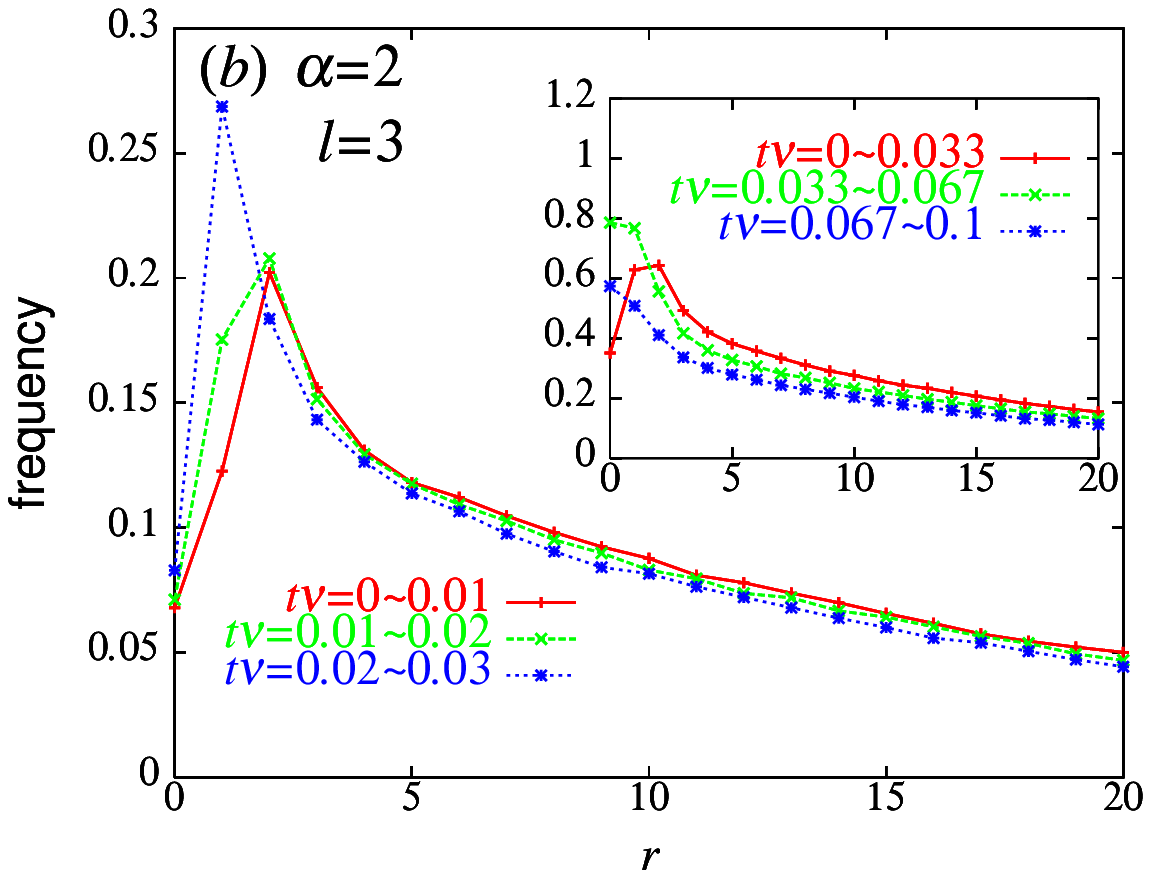}
\includegraphics[scale=0.65]{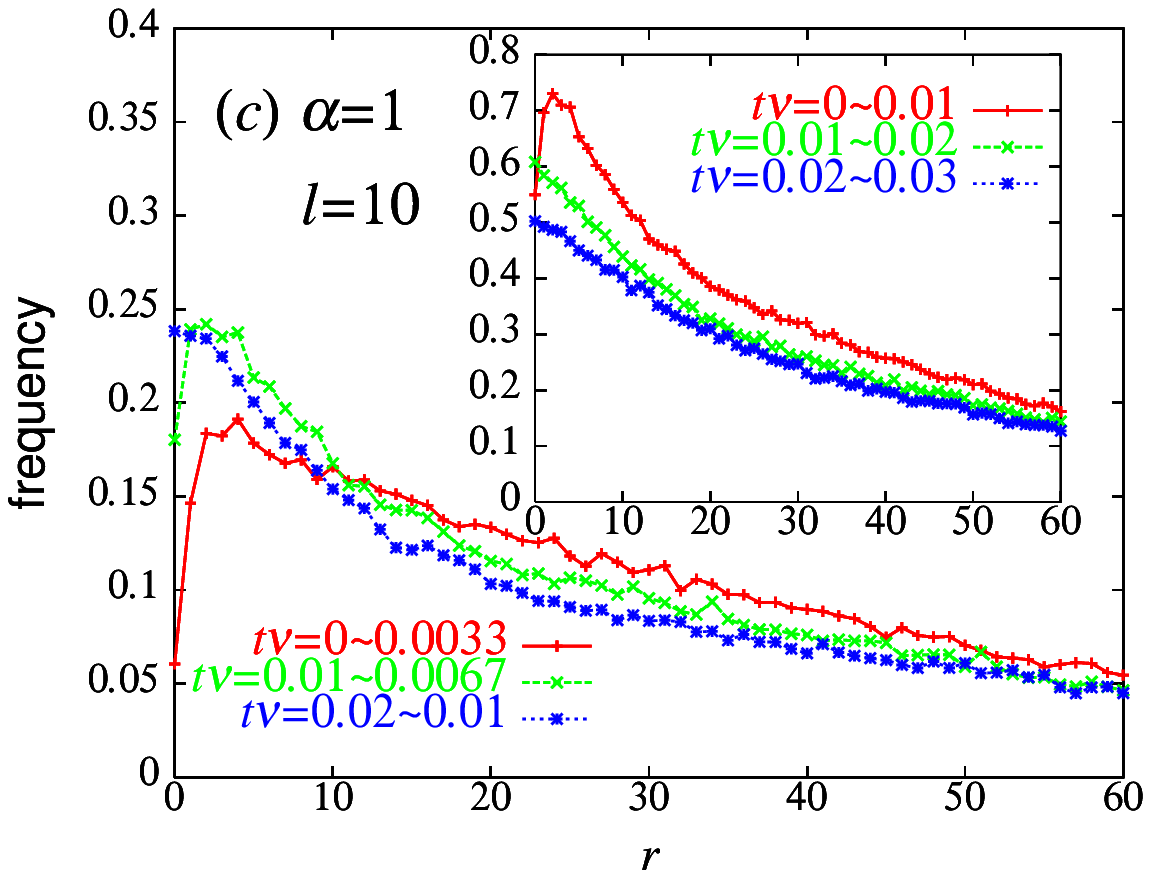}
\includegraphics[scale=0.65]{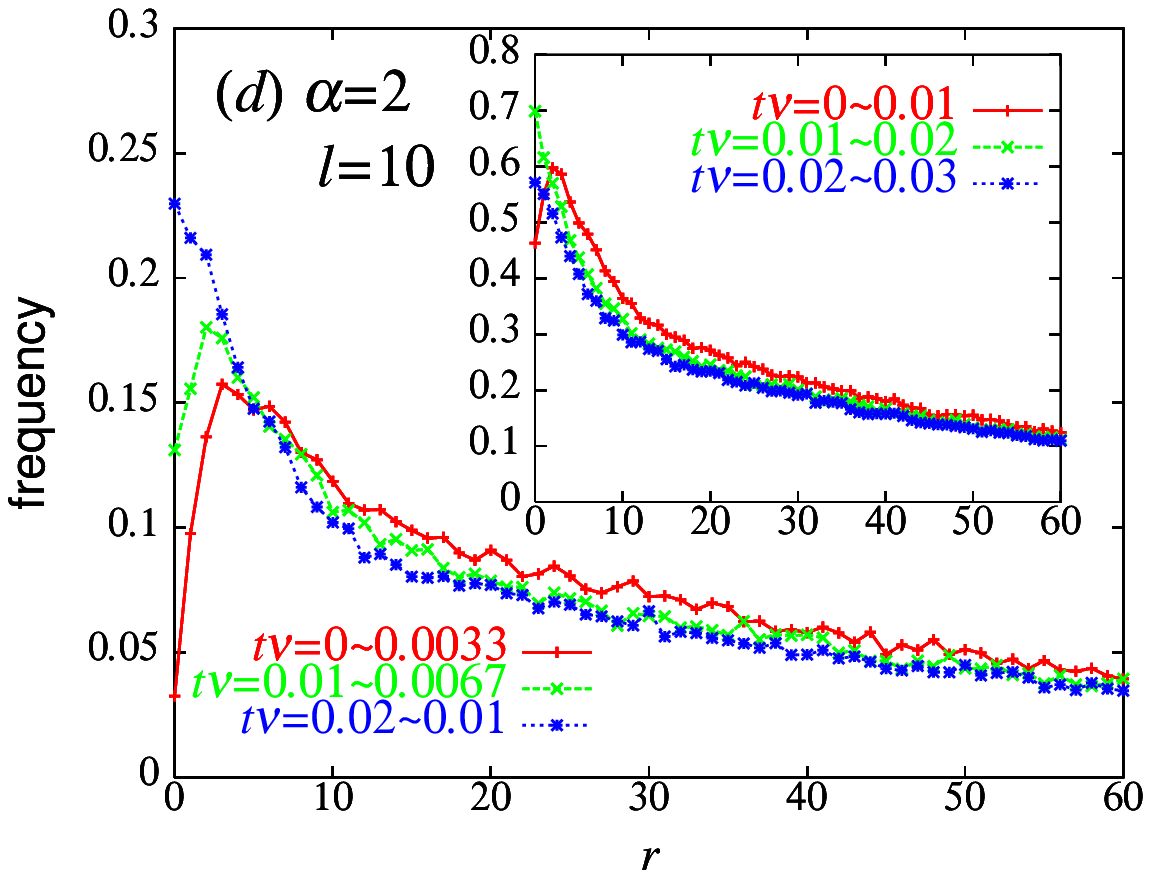}
\end{center}
\end{figure}
\begin{figure}
\caption{
Event frequency preceding the large event with $\mu >\mu_c=3$ plotted
versus $r$,
the distance from the epicenter of the upcoming mainshock, for several time periods before the mainshock. The parameters $\alpha$ and $l$ are 
$\alpha=1$, $l=3$ (a), $\alpha=2$, $l=3$ (b), $\alpha=1$, $l=10$ (c),
and $\alpha=2$, $l=10$ (d). The insets represent similar
plots with longer time intervals. The system size is $N=800$ in (a) and (b),
while it is $N=6400$ in (c) and (d).
}
\end{figure}

As can be seen from Fig.12, preceding the mainshock, 
there is a tendency
of the frequency of smaller events to be enhanced at and around 
the epicenter of the upcoming mainshock irrespective of the $\alpha$-value. 
For small enough $t$, such a cluster of smaller events correlated
with the large event may be regarded as foreshocks. 
Such an enhancement
of smaller events preceding the large event
was observed by \textit{Shaw et al.}, [1992].
Interestingly, however, as the mainshock becomes imminent, the frequency of
smaller events is {\it suppressed 
in a close vicinity of the epicenter 
of the
upcoming mainshock, though it continues to be enhanced in the surroundings\/}.
For real earthquake faults, such a quiescence phenomenon 
has been discussed as the ``Mogi doughnut'' [\textit{Mogi},
1969;\textit{Mogi}, 1979;\textit{Scholz},
1990]. 
The present observation contrasts with the earlier work 
of Carlson,
who claimed that the 1D BK model did not exhibit such a doughnut-like 
quiescence [\textit{Carlson}, 1991a]. 
We note that the quiescence observed here 
occurs only in a close vicinity of the 
epicenter of the mainshock, 
within one or two blocks from the epicenter, and only 
at a time close to the mainshock. 
The time scale for the appearance
of the doughnuts-like 
quiescence depends on the  $\alpha$ and $l$ values.
Namely, the time scale of the onset of the doughnut-like 
quiescence tends to be shorter for smaller
$\alpha$ and larger $l$, as can  be seen from Figs.12(a)-(d).

The doughnut-like quiescence observed here is closely related to the 
characteristic interval time scales associated with the single-block
event, $t_{int}^{(1)}$ and $t_{int}^{(2)}$, discussed in the previous 
subsection. Namely, the time scale of the onset of the doughnut-like
quiescence is basically determined by $t_{int}^{(1)}$ and $t_{int}^{(2)}$.
Indeed, the  time scale of the onset of the doughnut-like
quiescence turns out to be well described by the interval times
calculated in Fig.11 including its 
dependence on the parameters $\alpha$ and $l$.
For example, in the case of $\alpha=1$ and $l=3$ corresponding to Fig.12(a),
the interval 
time $t_{int}^{(1)}$ is estimated to be $t_{int}^{(1)}\nu =0.016$, 
which gives a reasonable measure of the onset time scale of the doughnut-like
quiescence observed in Fig.12(a).

In the present model, the size of the
``hole'' of the doughnut-like quiescence  as well as its 
onset time scale have no correlation with the
magnitude of the upcoming event, as has been expected  from its
interpretation in terms of $t_{int}^{(1)}$ and $t_{int}^{(2)}$. 
In other words, 
the doughnut-like quiescence is not peculiar to
large events in the present model. 
This means that, by monitoring the onset of the ``hole'' in the
seismic pattern, one can certainly deduce the time
and the position of the upcoming event, but unfortunately,
cannot tell about its magnitude. 
Yet, one might get some information about the magnitude of the upcoming event,
not from the size and the onset time of the ``hole'', but from the size
of the ``ring'' surrounding the ``hole''.
Thus,
we show in Figs.13 the  spatial correlation functions  before the mainshock
in the time range $0\leq t \nu \leq 0.001$ for the case of $\alpha=2$ 
and $l=3$, 
with varying the magnitude range of the upcoming
event. In the figure, 
the direction in which the rupture propagates farther in the upcoming event
is always taken to be the positive direction $r>0$, whereas the direction in which the rupture propagates less
is taken to be the negative direction $r<0$. As can be seen from
the figure, although the size of the ``hole'' around the origin $r=0$ has
no correlation with the magnitude of the upcoming event as mentioned above,
the size of the region of the active seismicity surrounding this ``hole'' is
well correlated with the size and the direction of the rupture of the
upcoming event. This coincidence might enable 
one to deduce the position and the size
of the upcoming event by monitoring the pattern of foreshocks, although
it is still difficult to give a pinpoint prediction of the time of the upcoming
mainshock. We note that such a correlation between the size of the seismically active region
and the magnitude of the upcoming event was observed in the BK model 
by \textit{Pepke et al}  [1994], and was examined in earthquake data as well
[\textit{Kossobokov and Carlson}, 1995].

\begin{figure}[ht]
\begin{center}
\includegraphics[scale=0.65]{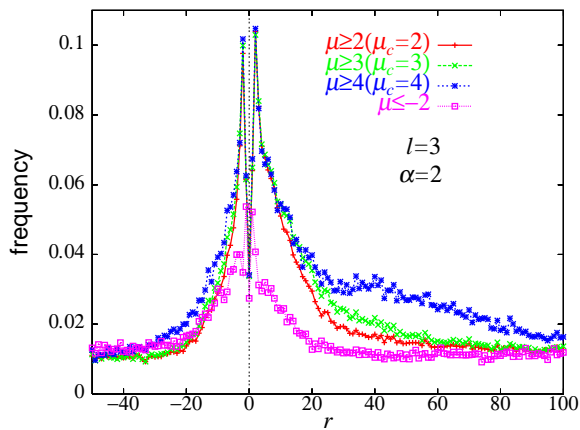}
\end{center}
\caption{
Event frequency preceding the events of various magnitude range
plotted versus $r$, 
the distance from the epicenter of the upcoming mainshock.
The curves correspond to the large events with $\mu >\mu_c=2,3$ and 4, and to
the smaller events with $\mu <-2$.
The direction in which the rupture propagates farther 
in the upcoming mainshock 
is always taken to be the positive direction $r>0$, whereas the direction in which the rupture propagates less
is taken to be the negative direction $r<0$.
The parameters are taken to be $\alpha=2$ and $l=3$, the time range being 
$0 \leq t \nu \leq 0.001$ before the mainshock. The system size is $N=800$.
}
\end{figure}

\subsection{Spatial correlation of events after the mainshock}

In this subsection, we examine the time-development of 
spatial correlations of events
{\it after\/} the mainshock.  The calculated spatial correlation functions
are shown in Figs.14 for various cases of the parameters, {\it i.e.\/},
$\alpha=1,\ l=3$ (a), $\alpha=2,\ l=3$ (b), $\alpha=1,\ l=10$ (c), and 
$\alpha=2,\ l=10$ (d). As can be seen from the figures,
after large events, 
small events remain active in a vicinity of the epicenter of
the mainshock which may be regarded as aftershocks. Events are suppressed in the surrounding region where
the displacement of the block was largest in the mainshock: See Fig.4. Hence,
it appears that
the seismicity after the large event in the present model
occurs so as to compensate
the rupture in the mainshock. However, the seismicity near the
epicenter is kept almost constant in time for some period after the mainshock, 
which is in apparent contrast to the power-law decay 
as expected from the Omori law:
See the insets of Figs.14. At longer time scales, the seismicity 
near the epicenter exhibits the non-trivial time dependence as shown in
the main panel of Figs.14. In the case of $\alpha=1$, while
the seismic activity near the
epicenter  seems to decay monotonically in this time range,  
the decay observed here is not a power-law decay as expected 
from the Omori-law.
Hence, aftershocks obeying the Omori-law is not realized in 
the BK model, as already reported [\textit{Carlson and Langer},
1989a;\textit{Carlson and Langer}, 1989b]. 
This is in apparent contrast to the observation for real faults.

Such an absence of 
aftershocks in the BK model might give a hint to the physical origin
of aftershocks obeying the Omori-law, {\it e.g.\/}, 
they may be driven by the slow chemical process 
at the fault, 
or by the elastoplaciticity associated with the ascenosphere, 
{\it etc\/}, which are  not taken into account in the present model.

\begin{figure}[ht]
\begin{center}
\includegraphics[scale=0.65]{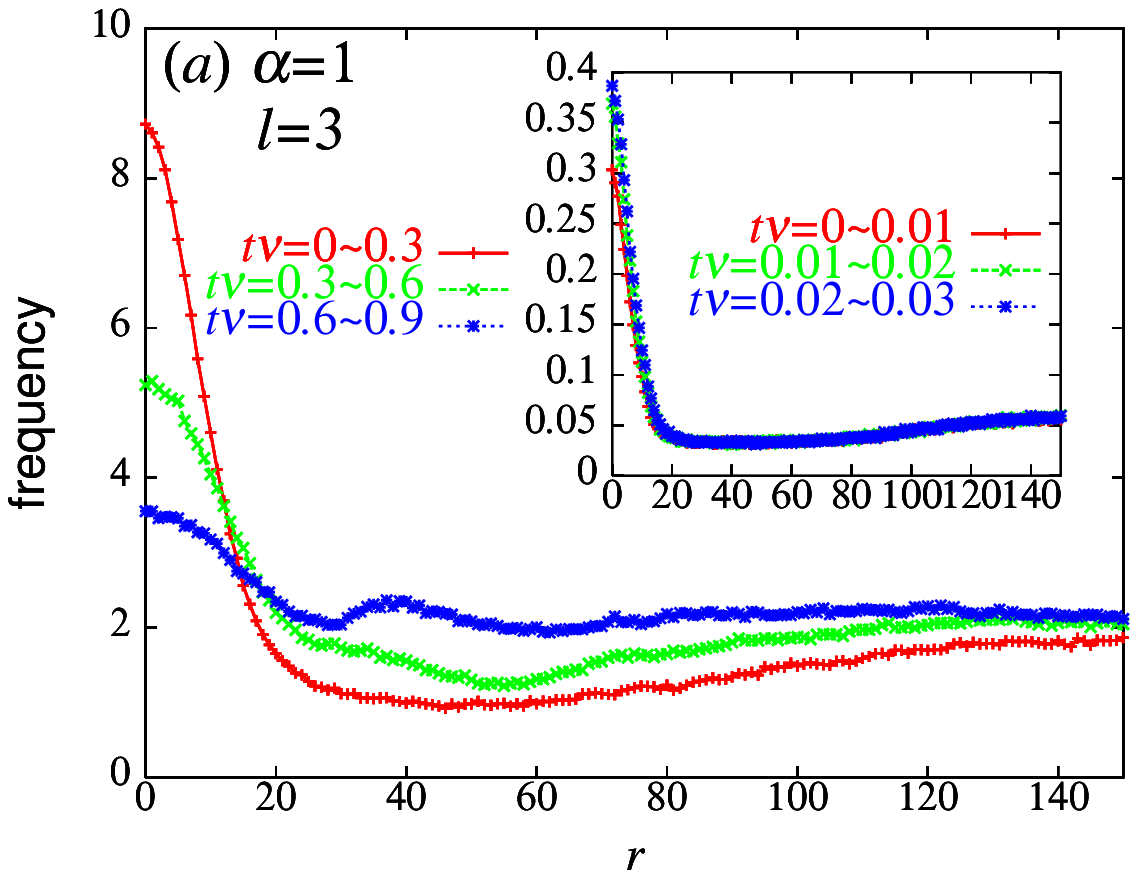}
\includegraphics[scale=0.65]{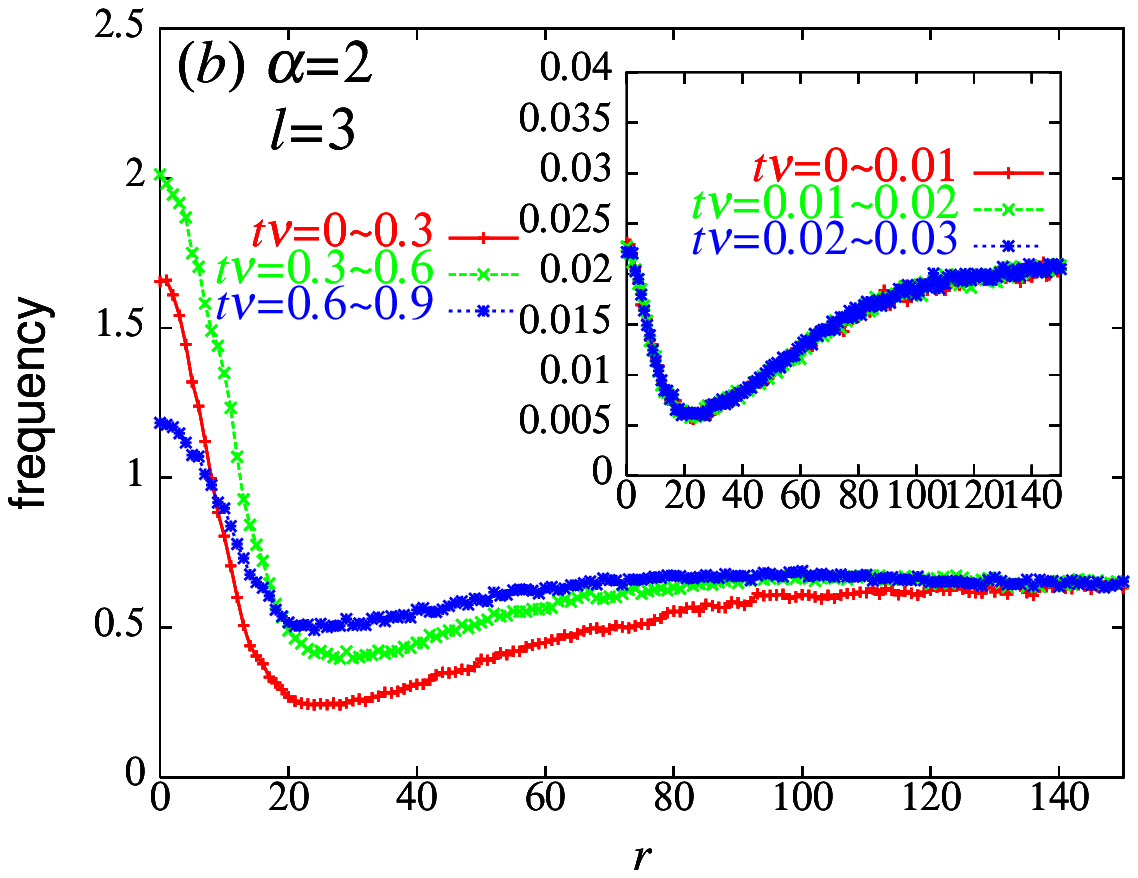}
\includegraphics[scale=0.65]{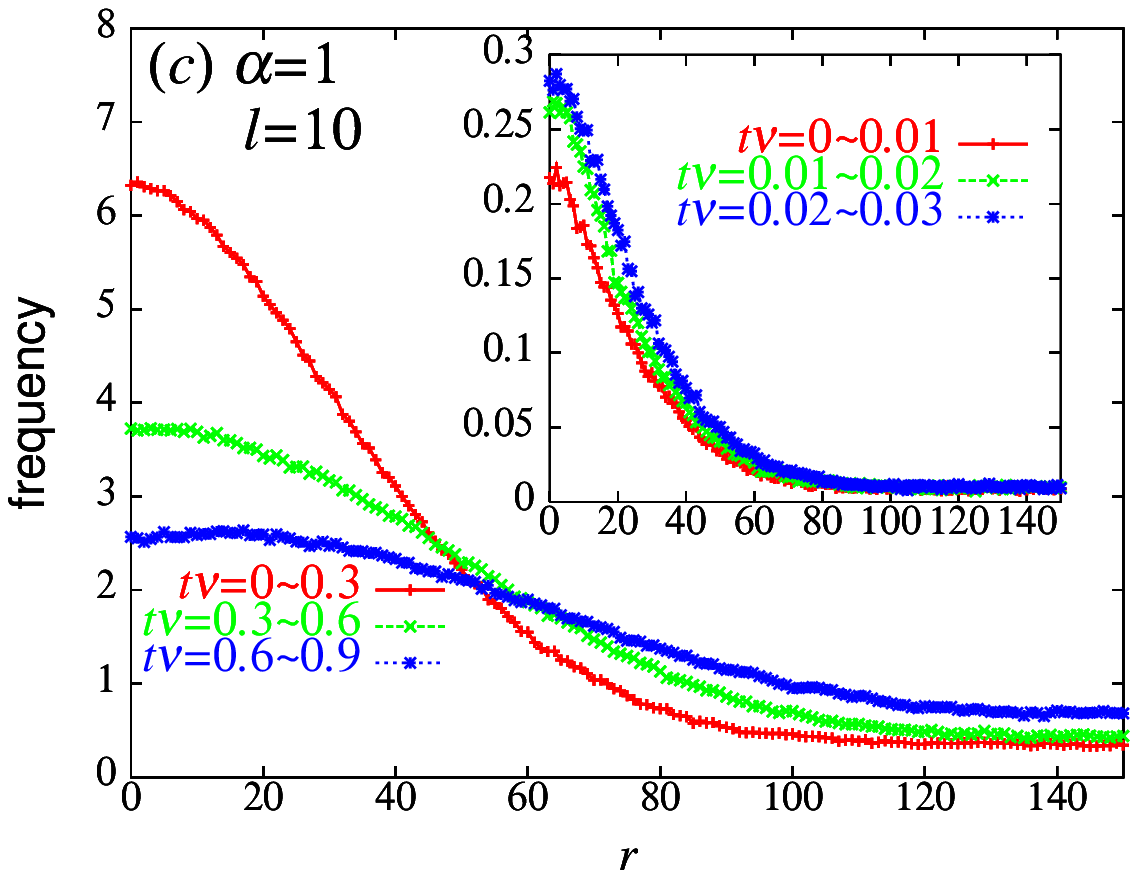}
\includegraphics[scale=0.65]{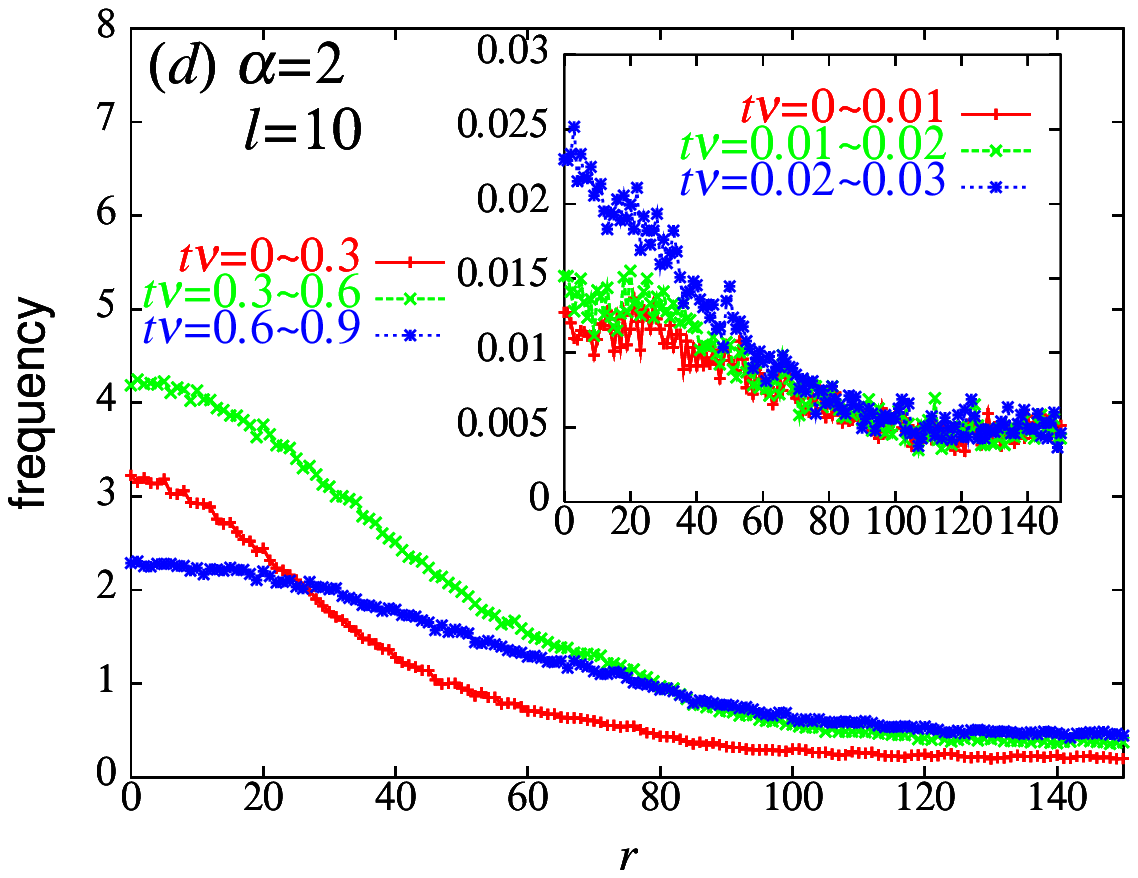}
\end{center}
\end{figure}
\begin{figure}
\caption{
Event frequency after the large event with $\mu >\mu_c=3$ plotted
versus $r$,
the distance from the epicenter of the upcoming mainshock, 
for several time periods after the mainshock. The parameters $\alpha$ and $l$ 
are 
$\alpha=1$, $l=3$ (a), $\alpha=2$, $l=3$ (b), $\alpha=1$, $l=10$ (c),
and $\alpha=2$, $l=10$ (d).
The main panel in each figure corresponds to the longer 
time scale, while the inset 
in  each figure corresponds to the shorter time scale. The system size is
$N=800$ in (a) and (b), while it is $N=6400$ in (c) and (d).
}
\end{figure}

\subsection{Time-dependent magnitude distribution}

\begin{figure}[ht]
\begin{center}
\includegraphics[scale=0.65]{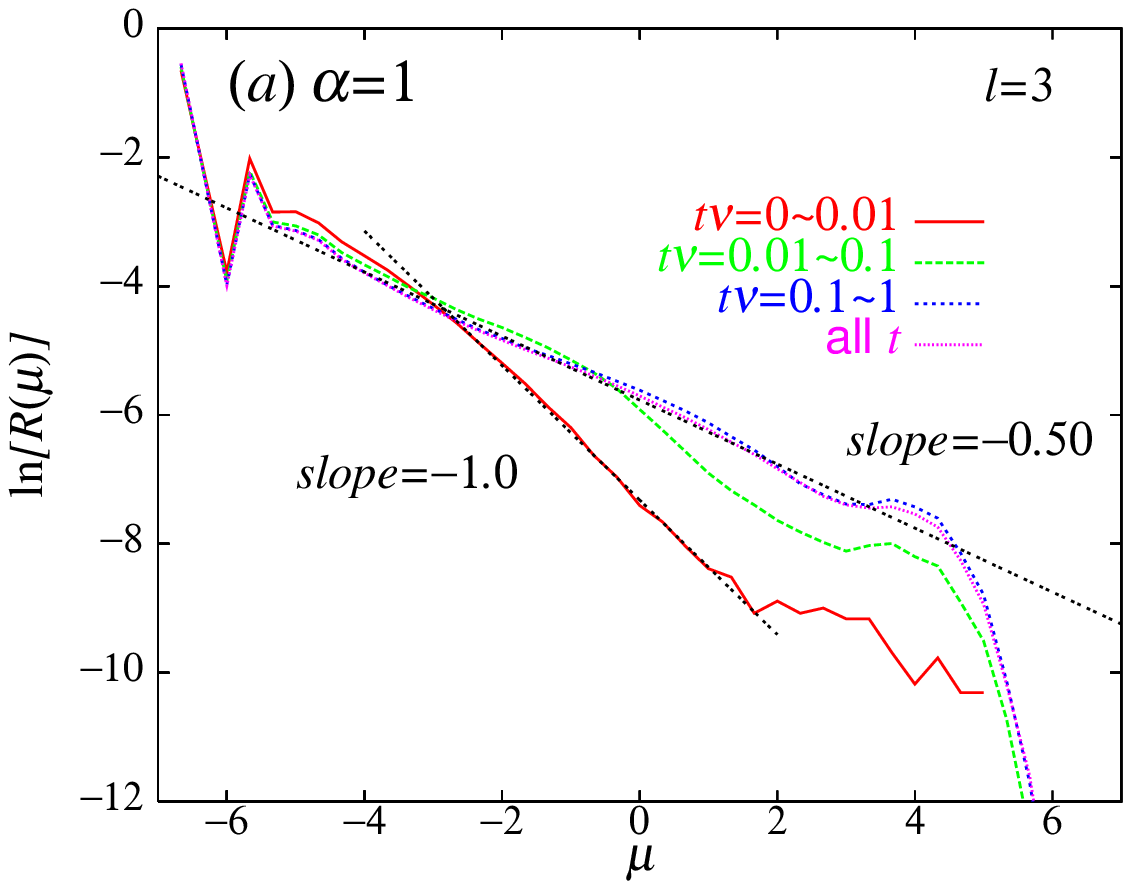}
\includegraphics[scale=0.65]{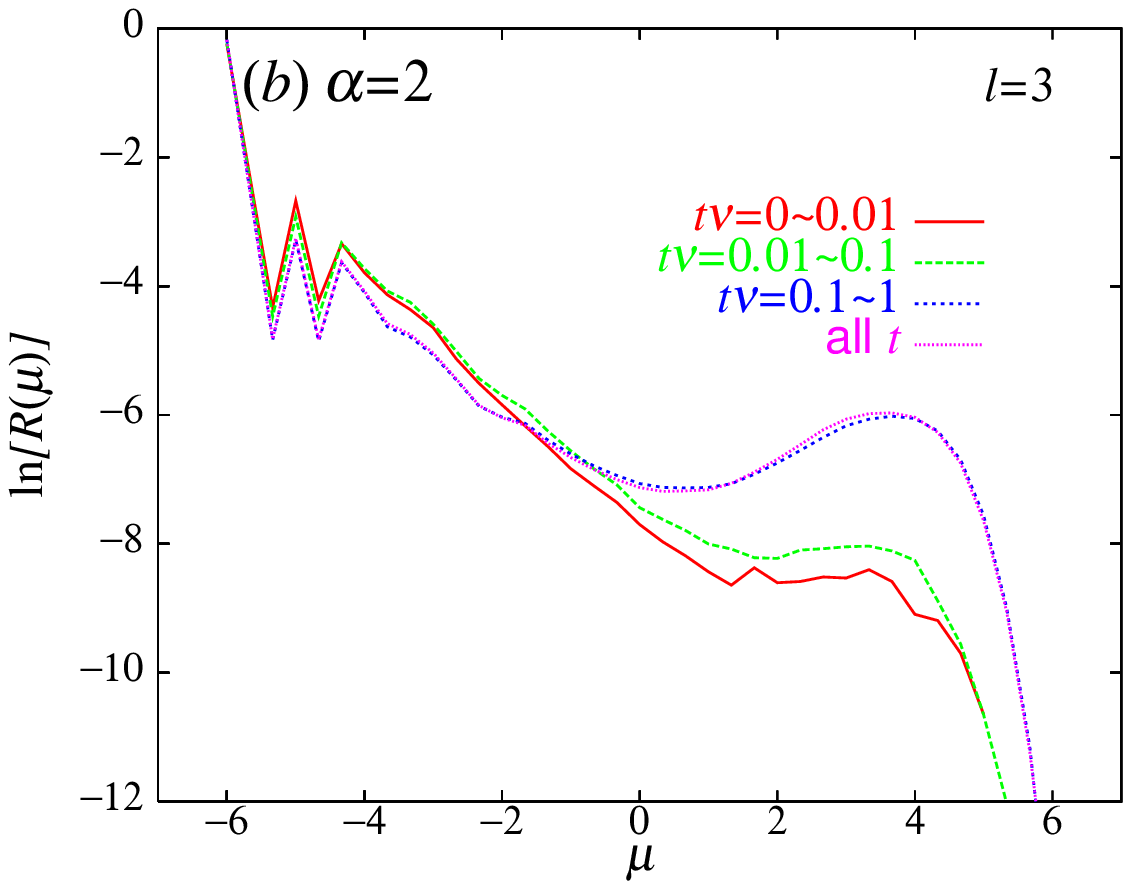}
\end{center}
\caption{
Local magnitude distribution before the mainshock with $\mu >\mu _c=3$ for several time periods before the mainshock for the cases of
$\alpha=1$ (a) and of $\alpha=2$ (b). The system size is $N=800$.
}
\end{figure}

As an other signature of the precursory phenomena,
we show in Fig.15 a ``time-resolved''  local magnitude distribution
for several time periods before the large
event, in the case of $\alpha =1$ (Fig.15(a)) and  $\alpha =2$ (Fig.15(b)). 
Only the events with their epicenters within 30 blocks from the
upcoming mainshock is counted here. While the parameter $l$ is held fixed to
$l=3$ here and in the following subsections, 
the qualitative trend turns out to be similar for other values of 
$l$.
As can be seen from the figures, as the mainshock approaches, 
the form of the magnitude distribution changes significantly. In particular, in the case of $\alpha=1$, 
the $B$-value describing the 
power-law regime tends to {\it increase\/}  as the mainshock
approaches, from the time-averaged mean value $\simeq 0.50$ to the value
$\simeq 1.0$ just before the mainshock: It is almost doubled.
Interestingly, a 
similar increase of the apparent $B$-value
preceding the mainshock was reported for some real faults 
[\textit{Smith}, 1981], while an opposite tendency, {\it i.e.\/},
a decrease of the apparent $B$-value preceding the mainshock,
was reported for other faults [\textit{Suehiro et al.}, 1964; \textit{Jaume and Sykes}, 1999]. 
For the case of larger $\alpha =2$ and 3, 
the change of the $B$-value preceding the mainshock is still appreciable,  
though in a less pronounced manner.

In Fig.16, we show a similar time-resolved  local magnitude distribution,
but now {\it after\/} the large
event, for the cases of $\alpha =1$ (Fig.16(a)) and $\alpha =2$ (Fig.16(b))
with fixing $l=3$.
As can be seen from the figure, 
the form of the magnitude distribution changes only very little after the mainshock,
in contrast to the one before the mainshock.

\begin{figure}[ht]
\begin{center}
\includegraphics[scale=0.65]{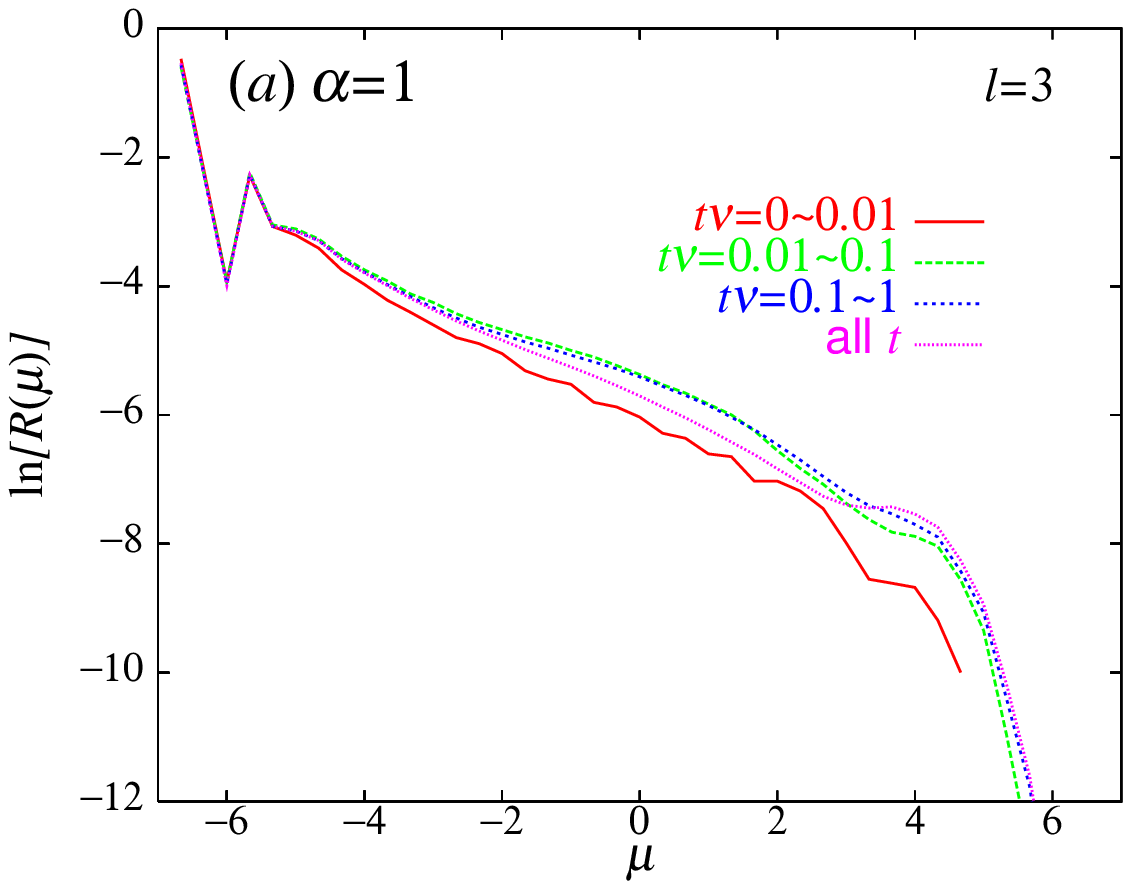}
\includegraphics[scale=0.65]{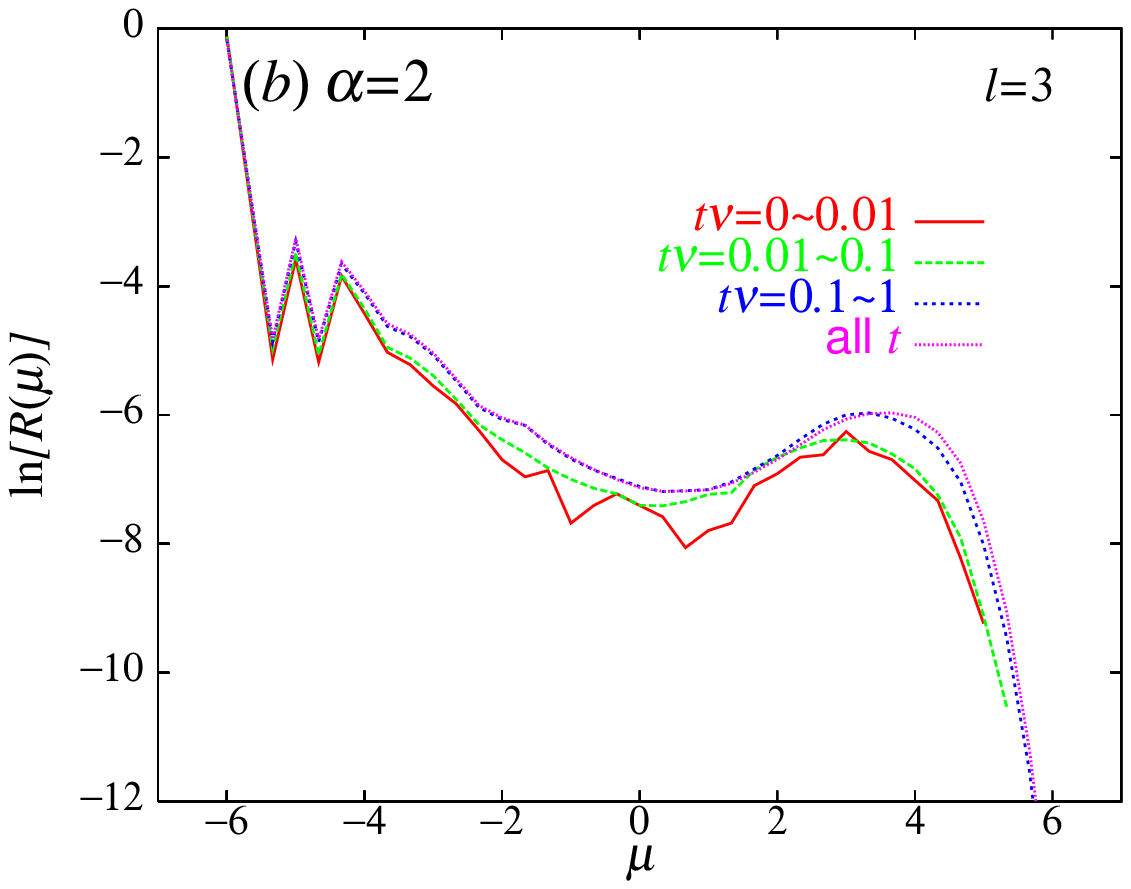}
\end{center}
\caption{
Local magnitude distribution after the mainshock with $\mu >\mu _c=3$ for several time periods after the mainshock for the cases of
$\alpha=1$ (a) and of $\alpha=2$ (b). The system size is $N=800$.
}
\end{figure}

\subsection{Time-predictable model versus size-predictable model}

In describing the time sequence of earthquake events, two simplified models 
have  widely been used, {\it i.e.\/}, the size-predictable model and the 
time-predictable model [\textit{Shimazaki and
Nakata},1980;\textit{Scholz}, 1990]. 
These two models are considered to describe the 
two extreme situations of the earthquake occurrence.  

In the size-predictable model,
one assumes that the stress is released to some
constant level just  after the event, 
whereas the stress level at the onset of the rupture
is random and unpredictable, as illustrated in Fig.17(a). 
Within this scheme, 
when one can monitor the ongoing status of the stress,  
one can predict  the size of an earthquake if it occurs at any given time,
although the time of the next event is unpredictable. 

In the time-predictable model,  on the other hand, 
one assumes that an earthquake occurs whenever the accumulated 
stress
reaches some constant level, whereas the size of the earthquake, 
{\it i.e.\/}, the
extent of the stress drop at the earthquake, is random and 
unpredictable, as illustrated in Fig.17(b). 
One can
predict within this scheme the time of the next event, 
though the size of the event is unpredictable.

\begin{figure}[ht]
\begin{center}
\includegraphics[scale=0.65]{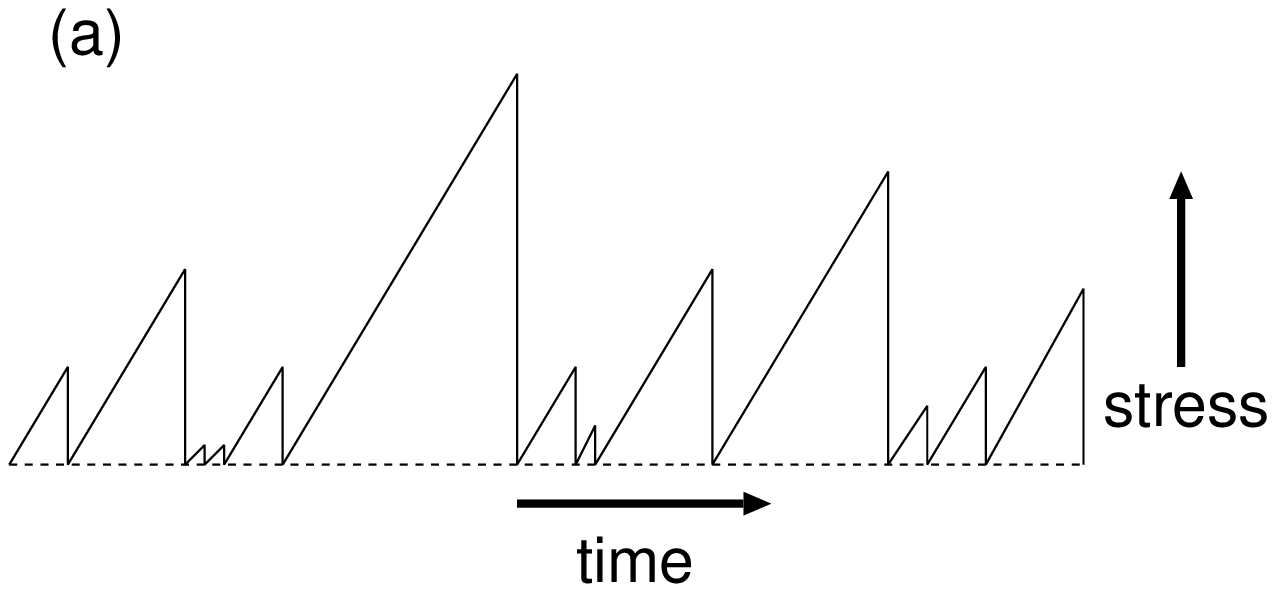}
\includegraphics[scale=0.65]{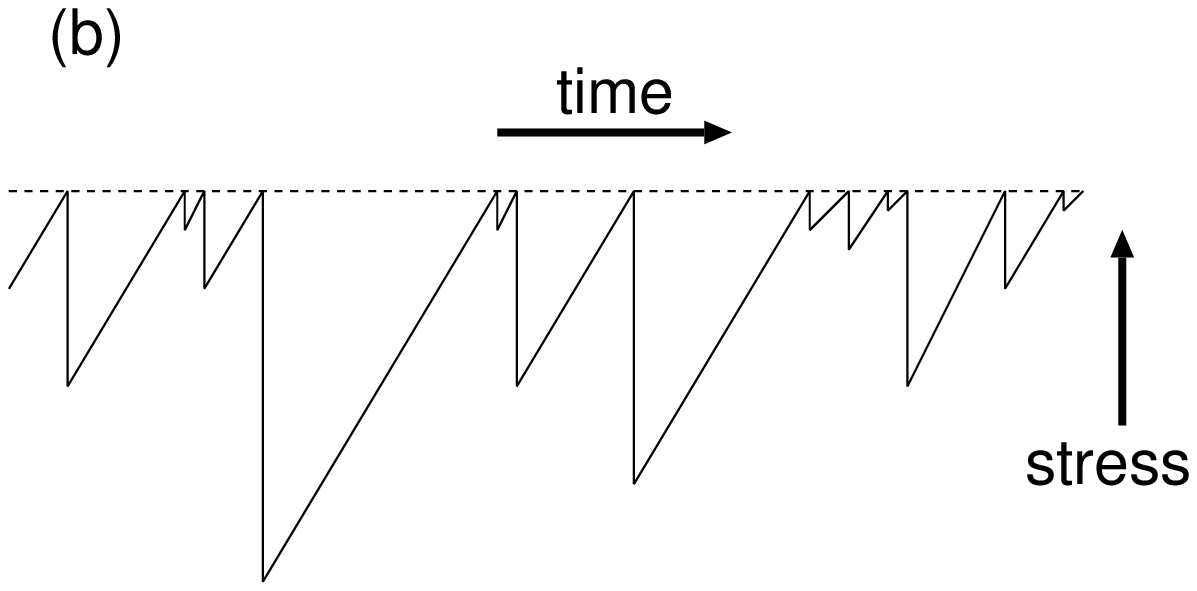}
\end{center}
\caption{
The size-predictable model (a) versus the time-predictable model (b).
}
\end{figure}

In this subsection, we examine whether 
the time sequence of large events (with $\mu \geq \mu_c=3$) 
of the present model 
is describable either by the time-predictable model or by the size-predictable 
model. Concerning the magnitudes of the successive earthquake events,
both models predict no correlation. Our analysis 
of Figs.8 and 9  above revealed that
the $\alpha=1$ model yielded no correlation between the magnitudes of the 
two successive events, while the $\alpha\geq 2$ models
yielded an 
anti-correlation between the magnitudes of the two successive events.
Hence, concerning the magnitude correlation of the successive events, 
the $\alpha=1$ model is consistent with  either the time-predictable
model or the size-predictable model, while the $\alpha\geq 2$ models are
consistent with neither of them.
In order to further examine the compatibility of the 
model with the size- and time-predictable models, we show in Fig.18
the magnitude distribution of large events with varying the
elapsed time since the last large event (waiting time). While the parameter $l$
is fixed to $l=3$ here, qualitative trend remains the same for other values of 
$l$.

When the size-predictable model applies, the next event tends to be larger as one waits longer till the next event, 
whereas, when the time-predictable model applies,
there should be no correlation
between  the waiting time and the size of the next event. As can be seen from 
Fig.18, there is a clear tendency irrespective
of the value of $\alpha$
that the next event tends to be larger as the waiting time gets longer. 
This property is compatible with the size-predictable model, 
but is incompatible 
with the time-predictable model.

In Fig.19, we show the distribution of the waiting time defined above 
with varying the 
magnitude of the last large event. When the time-predictable model applies,
the waiting time tends to get longer as the size of the last large event
gets larger,
whereas,  when the 
size-predictable model applies, there should be no correlation
between the size of the last large event and the waiting time.
As can be seen from Fig.19, the $\alpha=1$ model exhibits no correlation between
the waiting time and the size of the last event, which
is consistent with the size-predictable model, whereas the $\alpha\geq 2$ 
models
exhibit a positive correlation between  the waiting time and the size of the last event, which is 
apparently consistent with the time-predictable model.

In view of all these observations, one may conclude that the $\alpha=1$
BK model is well describable by the size-predictable model, whereas the
$\alpha=2,3$ BK models are describable by neither the time-predictable nor
the size-predictable model.

\begin{figure}[ht]
\begin{center}
\includegraphics[scale=0.65]{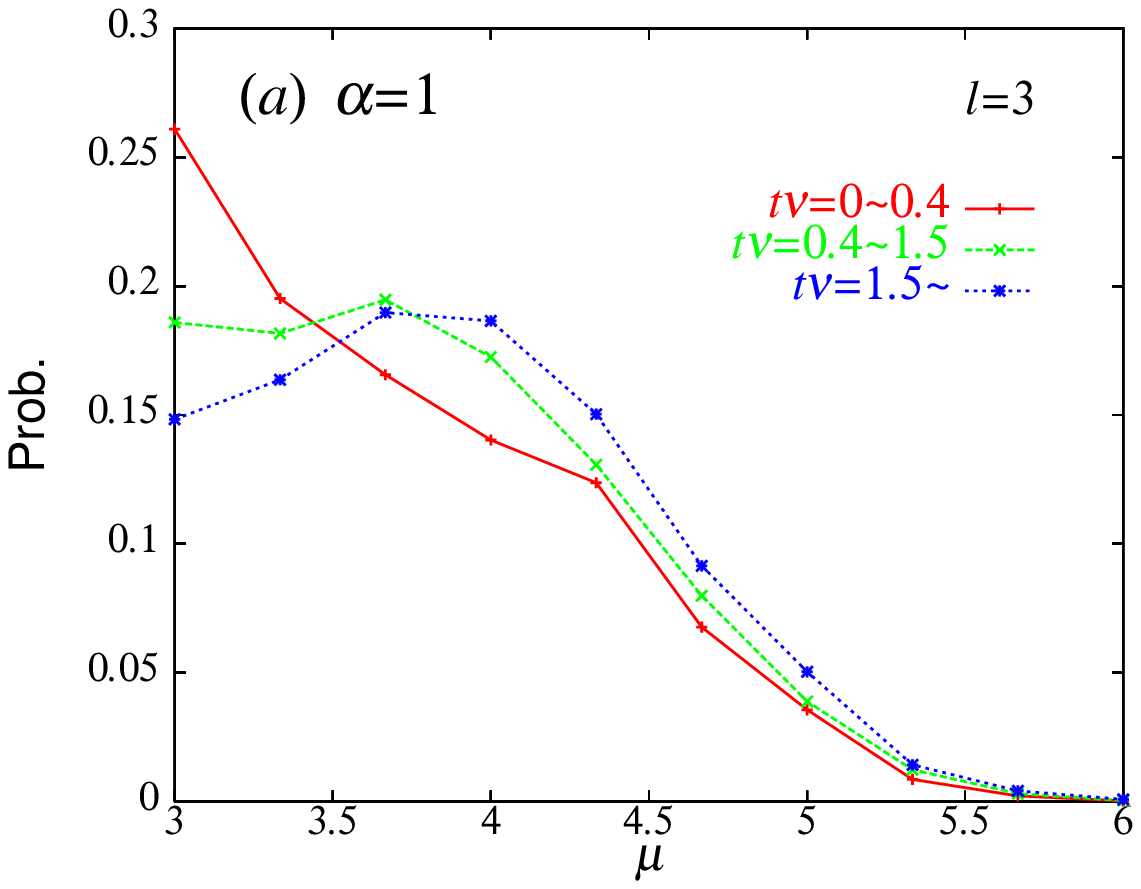}
\includegraphics[scale=0.65]{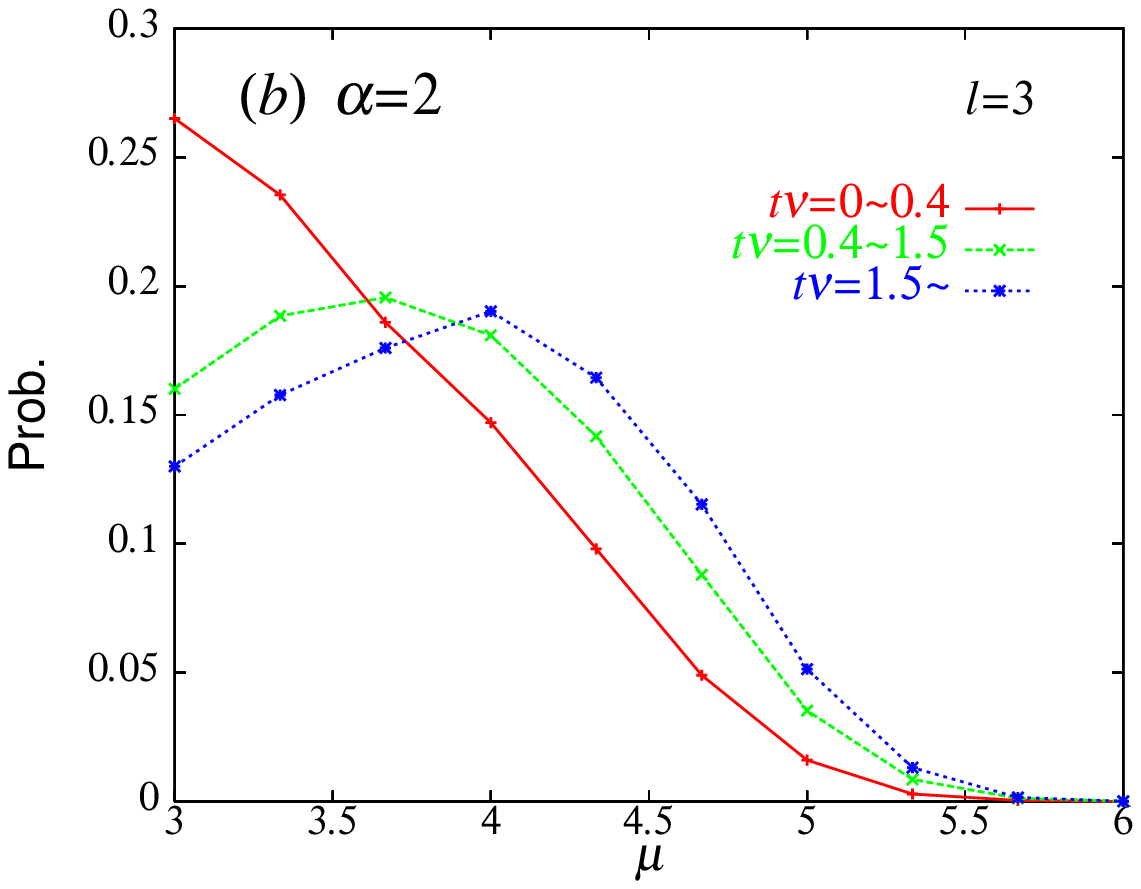}
\end{center}
\caption{
The magnitude distribution of large events with $\mu\geq \mu_c=3$ for various values of the waiting time, {\it i.e.\/}, the time elapsed since the last large event with $\mu\geq \mu_c=3$ which occurred within 30 blocks from the epicenter of the event. The parameter $\alpha$ is $\alpha=1$ (a) and $\alpha=2$ (b), whereas the parameter $l$ is fixed to be $l=3$. The system size is $N=800$.
}
\end{figure}

\begin{figure}[ht]
\begin{center}
\includegraphics[scale=0.65]{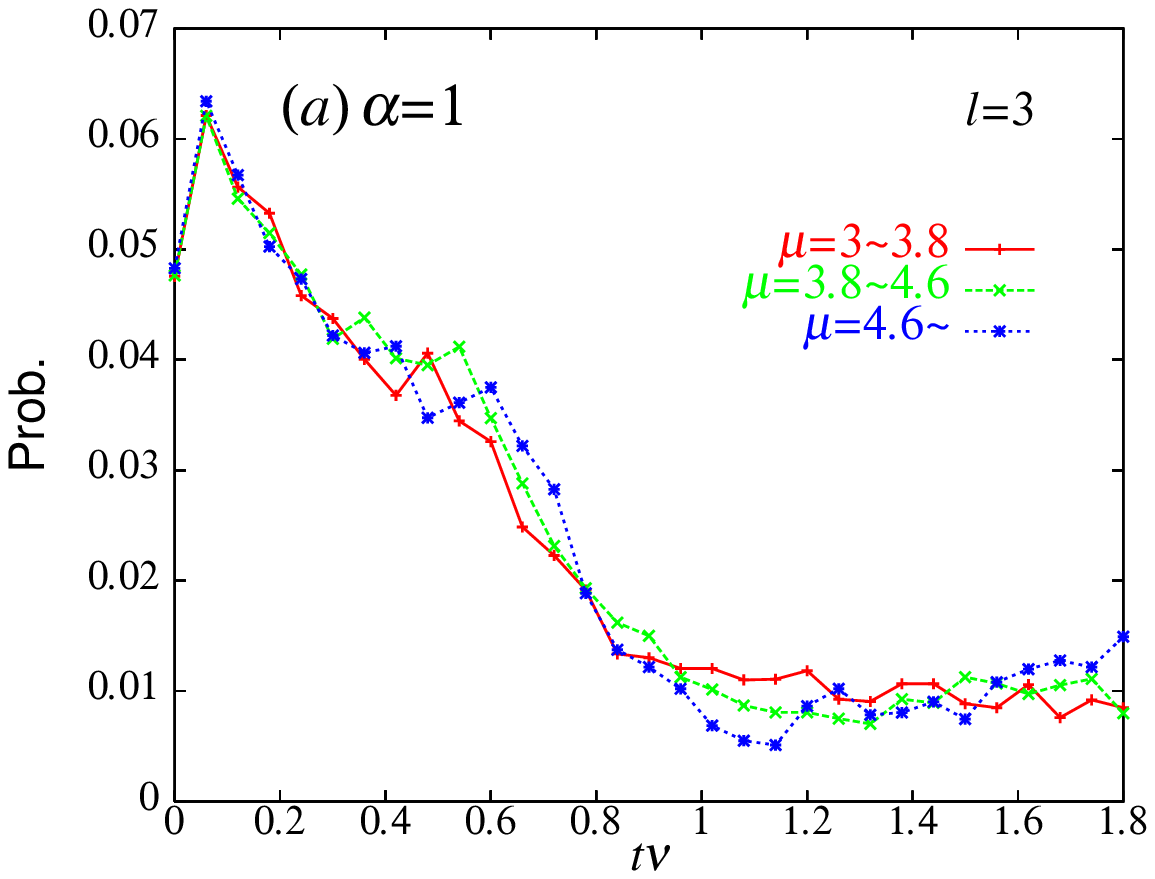}
\includegraphics[scale=0.65]{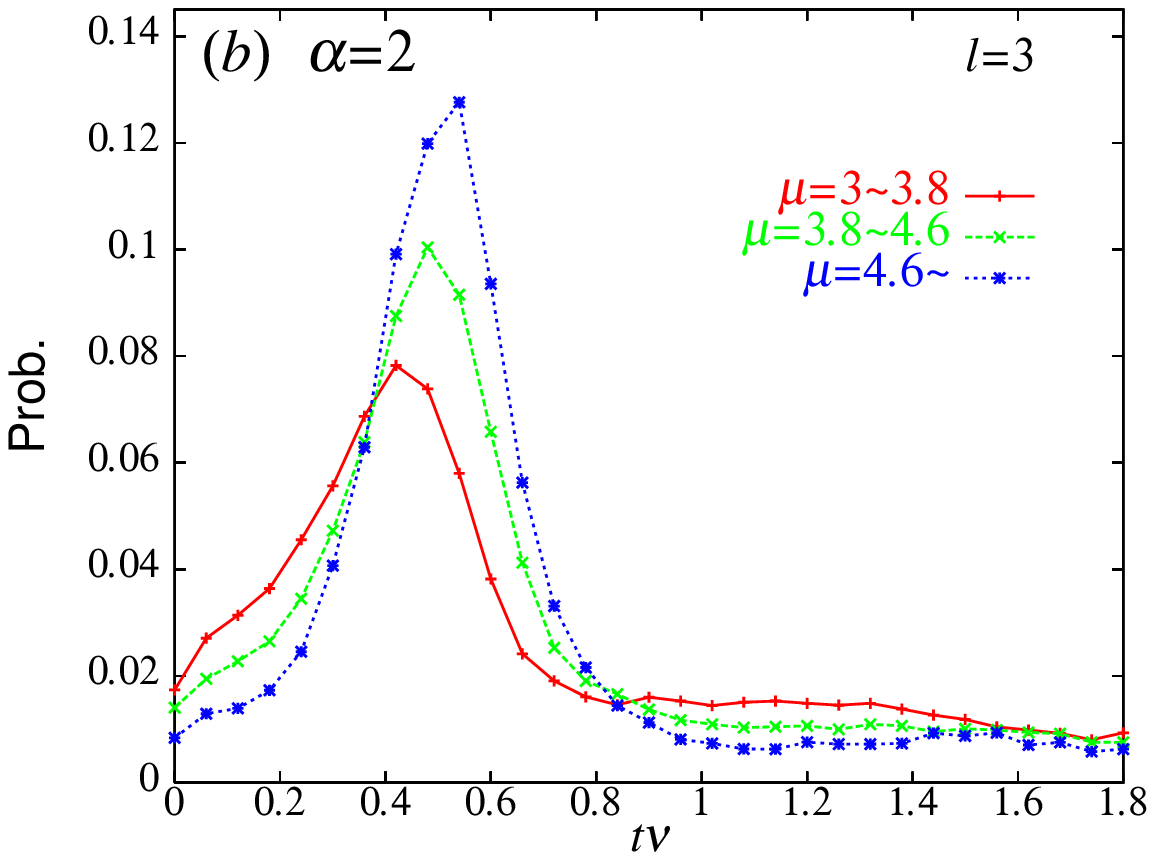}
\end{center}
\caption{
The distribution of the waiting time, {\it i.e.\/}, the time elapsed since the last large event with $\mu\geq \mu_c=3$ which occurred within 30 blocks from the epicenter of that event,  for various values of the magnitude of the last large event. The parameter $\alpha$ is $\alpha=1$ (a) and $\alpha=2$ (b), whereas the parameter $l$ is fixed to be $l=3$. The system size is $N=800$.
}
\end{figure}

\section{Summary and discussion}

In summary, we studied the spatio-temporal correlations of the 1D BK model of
earthquakes with main focus on how the statistical properties of earthquakes
depend on
the frictional and elastic properties of earthquake faults. We have found that when the extent 
of the velocity-weakening 
property gets smaller, the system tends to be more critical, while, as the
velocity-weakening property is enhanced, the system tends to be more 
off-critical with enhanced features of characteristic earthquakes.
Periodic feature of large events 
is eminent when
the friction force exhibits a strong frictional instability, whereas
when the friction force exhibits a weak frictional instability 
large events often occur as twin and/or unilateral events. The model with 
weaker frictional instability, {\it i.e.\/}, the $\alpha=1$ model, is well describable by the size-predictable model, while the model with stronger frictional instability, {\it i.e.\/}, the $\alpha \geq 2$  
models, are  describable neither by the size-predictable model nor the time-predictable model.
We also observed several intriguing precursory phenomena associated with large events.
Preceding the mainshock, the frequency of smaller 
events is gradually
enhanced, whereas just before the mainshock it is 
suppressed only in a close vicinity of the epicenter of the upcoming mainshock
(the Mogi doughnut). On the other hand, the Omori law of aftershocks is not observed in the present model. When the friction force exhibits a weak frictional 
instability, 
the apparent $B$-value of the magnitude distribution increases significantly
preceding the mainshock. 

We note that the statistical properties analyzed here 
often depend sensitively on the
parameter $\alpha$, but not depend much on the parameter $l$, at least at
qualitative level, so long as one deals with sufficiently large systems. 
Some of the statistical properties, {\it e.g.\/}, the global recurrence-time
distribution, are subject to significant finite-size effects particularly 
when the parameter $l$ is large. In such cases, care has to be taken in
separating the intrinsic bulk property of the model 
from the finite-size property.

Of course, earthquake is a complex phenomenon, and it is not a trivial matter at all how faithfully the statistical
properties as observed here for the 1D BK model represent those of
real earthquakes. After all, the BK model is a highly simplified statistical
model
where many features of real fault system have been neglected or simplified.
Still, the model is expected to serve as a useful reference
in identifying and elucidating the nature of real earthquakes. In particular,
some of the precursory phenomena as observed here may have
relevance to real earthquakes. We hope that the present study might give a
step toward the fuller understanding of the statistical properties of earthquakes as a stick-slip dynamical instability.

\end{document}